\documentclass[twocolumn,Astrosymb]{aastex63}

\newcommand{\kms} {\rm{km~s}$^{-1}$} 
\newcommand{\ci} {\rm{cm}$^{-1}$}  
\newcommand{\csi} {\rm{cm}$^{-2}$}  
\newcommand{\cci} {\rm{cm}$^{-3}$}  
\newcommand{\si} {\rm{s}$^{-1}$}  

\newcommand{\hii}{\rm{H\textsc{ii}}}
\newcommand{\acet} {$\mathrm{C}_{2}\mathrm{H}_{2}$}

\newcommand{\htwo} {$\mathrm{H}_{2}$}
\newcommand{\nhtwo}{$N_{\mathrm{H}_2}$}
\newcommand{\water}{$\mathrm{H}_{2}\mathrm{O}$}
\newcommand{\hcniso}{$\mathrm{H}^{13}\mathrm{CN}$}

\newcommand{\vfwhm}{$v_{\mathrm{FWHM}}$}
\newcommand{\vlsr}{$v_{\mathrm{LSR}}$}

\shorttitle{First MIR Detection of HNC}
\shortauthors{Nickerson et al.}
\begin{document}

\title{The First Mid-Infrared Detection of HNC in the Interstellar Medium: Probing the Extreme Environment Towards the Orion Hot Core}

\correspondingauthor{Sarah Nickerson}
\email{sarah.nickerson@nasa.gov}

\author[0000-0002-7489-3142]{Sarah Nickerson}
\affiliation{Space Science and Astrobiology Division, NASA Ames Research Center, Moffet Field, CA, 94035 USA}
\affiliation{Bay Area Environmental Research Institute, Moffet Field, CA, 94035, USA}

\author[0000-0001-9920-7391]{Naseem Rangwala}
\affiliation{Space Science and Astrobiology Division, NASA Ames Research Center, Moffet Field, CA, 94035 USA}

\author{Sean Colgan}
\affiliation{Space Science and Astrobiology Division, NASA Ames Research Center, Moffet Field, CA, 94035 USA}

\author[0000-0002-6528-3836]{Curtis DeWitt}
\affiliation{USRA, SOFIA, NASA Ames Research Center MS 232-11, Moffett Field, CA 94035, USA}

\author[0000-0003-2458-5050]{Xinchuan Huang}
\affiliation{Space Science and Astrobiology Division, NASA Ames Research Center, Moffet Field, CA, 94035 USA}

\author[0000-0002-0603-8777]{Kinsuk Acharyya}
\affiliation{Physical Research Laboratory: Ahmedabad, Gujarat, India}

\author[0000-0001-7479-4948]{Maria Drozdovskaya}
\affiliation{Center for Space and Habitability, University of Bern, Gesellschaftsstrasse 6, CH-3012 Bern, Switzerland}

\author[0000-0003-4716-8225]{Ryan C. Fortenberry}
\affiliation{Department of Chemistry and Biochemistry, University of Mississippi, 38655 USA}

\author[0000-0002-4649-2536]{Eric Herbst}
\affiliation{Departments of Chemistry and Astronomy, University of Virginia, McCormick Rd, Charlottesville, VA, 22904 USA}

\author[0000-0002-2598-2237]{Timothy J. Lee}
\affiliation{Space Science and Astrobiology Division, NASA Ames Research Center, Moffet Field, CA, 94035 USA}

\begin{abstract}
We present the first mid-infrared (MIR) detections of HNC and \hcniso\ in the interstellar medium, and numerous, resolved HCN rovibrational transitions. Our observations span 12.8 to 22.9 \micron\ towards the hot core Orion IRc2,  obtained with the Echelon-Cross-Echelle Spectrograph aboard the Stratospheric Observatory for Infrared Astronomy (SOFIA). Exceptional, $\sim5$ \kms\, resolution distinguishes individual rovibrational transitions of the HNC and HCN  P, Q, and R branches; and the \hcniso\ R branch. This allows direct measurement of the species' excitation temperatures, column densities, and relative abundances. HNC and \hcniso\ exhibit a local standard rest velocity of $-7$ \kms\ that may be associated with an outflow from nearby radio source I and an excitation temperature of about 100 K. We resolve two velocity components for HCN, the primary component also being at $ -7$ \kms\ with temperature 165 K. The hottest component, which had never before been observed, is at 1 \kms\ with temperature 309 K. This is the closest component to the hot core's centre measured to date.  The derived $^{12}$C/$^{13}$C=$13\pm2$ is below expectation for Orion's Galactocentric distance, but the derived HCN/HNC=$72\pm7$ is expected for this extreme environment. Compared to previous sub-mm and mm observations, our SOFIA line survey of this region shows that the resolved MIR molecular transitions are probing a distinct physical component and isolating the chemistry closest to the hot core.
\vspace{10mm}

\end{abstract}

\section{Introduction} \label{sec:intro}

The isomers hydrogen cyanide (HCN) and hydrogen isocyanide (HNC) have shown their ubiquity through a range of astrophysical phenomena since their first detection in Galactic star-forming regions \citep{Snyder1971,Zuckerman1972,Snyder1972,McGuire2018}. Galactic sources include comets in our Solar System \citep{Lis1997,Agundez2014}, Titan's atmosphere \citep{Moreno2011}, dark molecular clouds \citep{Irvine1984,Hirota1998}, diffuse clouds \citep{Turner1997, Liszt2001}, star-forming regions \citep{Loughnane2012,Tennekes2006}, protoplanetary objects \citep{Dutrey1997,Kastner1997,Hrivnak2000,Herpin2000,Graninger2015}, circumstellar envelopes \citep{Bujarrabal1994,Cernicharo1996,Cernicharo2013}, carbon stars \citep{Harris2003} and the Galactic centre's circumnuclear disk \citep{Harada2015}. The isotopes of HCN and HNC are useful for measuring nitrogen, carbon, and hydrogen isotopic ratios in star-forming regions and cores \citep{Wampfler2014,Zeng2017,Colzi2018a,Colzi2018b}. They also play a role in grain-surface chemistry at cold temperatures \citep{Lo2015}. HCN and HNC have also been detected in several external galaxies \citep{Rickard1977,Henkel1988,Gao2004b} including Seyfert galaxies \citep{Perez-Beaupuits2007}, molecular outflows \citep{Aalto2012}, and high redshift galaxies \citep{Guelin2007}.

Several studies have also discovered HCN or HNC in hot cores \citep{Goldsmith1981,Schilke1992,Lahuis2000,Knez2001,Boonman2001,Lacy2002,Knez2009,Rolffs2011}. These warm ($\ge 100$ K), dense ($10^5$ to  $10^8$ \cci) regions of the interstellar medium (ISM) near young, high-mass protostars. Stellar radiation heats the gas and dust grains; evaporates the icy dust mantles in the cold molecular cloud where the protostar formed; and reveals a chemically rich reservoir of complex organic molecules \citep{Ohishi1997,VanDerTak2004,Bisschop2007,Belloche2013,Rivilla2017}. Hot cores are possibly the antecedents to later ultra compact \hii\ regions \citep{Cesaroni2005}, representing a key stage in stellar evolution.  Similar hot corinos envelop low-mass protostars \citep{Bottinelli2004}.

The HCN/HNC abundance ratio changes under different conditions in star-forming regions, providing a useful probe of regional properties. HCN/HNC nears unity at low temperatures \citep{Irvine1984,Schilke1992}, infrared dark clouds \citep{Vasyunina2011,Miettinen2014} and dark cores \citep{Hirota1998}. Dark cores are defined as regions with temperatures $\sim$ 10 K and densities of $10^3$ to $10^5$ \cci\ \citep{Benson1989}. By comparison, in the Orion Molecular Cloud (OMC-1) region, the HCN abundance remains similar to these dark cores while the HNC abundance drops. HCN/HNC is $\sim$ 80 towards the Orion hot core IRc2, and drops to $\sim$ 5 away from the hot core \citep{Schilke1992}. Later observations in this region found a correlation between gas kinetic temperature and the HCN-to-HNC intensity ratio \citep{Hacar2020}. 

These isomers are a promising chemical clock. \citet{Jin2015} measured the HCN/HNC abundance ratio in massive star-forming regions at differing evolutionary stages (infrared dark clouds, high-mass protostellar objects, and ultra-compact \hii\ regions), revealing that this ratio increases towards more advanced stages. Indeed, the evidence suggests that HNC is more prone to destruction at higher temperatures than HCN, and will be less abundant in more advanced protostellar objects as they heat the ISM.

Theoretical calculations predict the isomers' main formation pathway to be the dissociative recombination of HCNH$^+$ with an electron \citep{Herbst1978}, and experiments find this produces near equal quantities of HCN and HNC \citep{Mendes2012}. However, HNC is the the less stable of the two isomers \citep{Lee1991,Bowman1993,Nguyen2015} and should be even less abundant than observed \citep[for an overview see][]{Loison2014}. Mechanisms proposed to regulate the HCN/HNC ratio include the gas-phase H + HNC reaction barrier \citep{Graninger2014}, UV dissociation \citep{Chenel2016,Aguado2017}, and collisions with \htwo\ \citep{HernandezVera2017} and He \citep{Sarrasin2010}.

Despite being the first hot core discovered \citep{Ho1979}, Orion IRc2 is atypical. Most hot cores are mainly internally heated, enveloping their protostar, while IRc2 is heated externally without evidence of any internal source \citep{Blake1996,Okumura2011}. A previous explosive outflow may have heated it while it was a preexisting dense clump, it may be itself an explosive outflow, or it is possibly a cavity offering a glimpse into a dust obscured protostar \citep{Shuping2004,Zapata2011,Goddi2011,Bally2017,Orozco-Aguilera2017}. The main candidate for IRc2's heat source is radio source I, a possible binary system of protostars \citep{Hirota2017}, which is heavily dust-obscured \citep{Plambeck2016} and considered to be the main energy source of the region \citep{Hirota2015}. Nonetheless IRc2's location in the Orion Molecular Cloud, the nearest massive star-forming region to Earth at $388\pm5$ pc \citep{Kounkel2017}, makes it ideal for observing the rich hot core chemistry.

Numerous HCN and HNC emission lines have been observed towards IRc2 in the mm \citep{Goldsmith1986,Schilke1992} and sub-mm \citep{Stutzki1988,Harris1995,Schilke2001,Comito2005}, two well-studied spectral regions due to to their accessibility from the ground. However, observations in the mid-infrared (MIR) have been scarce due to atmospheric absorption. The MIR is nonetheless critical for understanding fully the chemistry of the ISM. Additionally, rovibrational transitions for molecules with no permanent dipole moment are accessible only in the this wavelength range.

The Short-Wavelength Spectrometer aboard the \textit{Infrared Space Observatory} covered 2.38 to 45.2 \micron\ \citep{deGraauw1996}, but detection was limited to the strongest absorption features and not individual rovibrational transitions in hot cores \citep{VanDishoeck1998,Lahuis2000,Boonman2003}. Similarly, \textit{Spitzer}'s Infrared Spectrograph \citep{Houck2004} could only detect the strongest HCN absorption feature towards young stellar objects \citep{An2009,Lahuis2006,An2011}. The ground-based Texas Echelon Cross Echelle Spectrograph \citep[TEXES,][]{Lacy2002} resolves these individual transitions from 5 to 25 \micron\ with a maximum resolving power of $\sim 100,000$, but TEXES cannot access the entire MIR range due to atmospheric interference. HCN rovibrational and rotational transitions were detected with TEXES towards the hot cores AFGL 2591 \citep{Knez2001}, NGC 7538 IRS 1 \citep{Knez2009}, and IRc2 \citep{Lacy2002,Lacy2005}. 

The Stratospheric Observatory for Infrared Astronomy \citep[SOFIA,][]{Young2012} is an airbourne observatory that flies above 99\% of atmospheric water vapour. The Echelon-Cross Echelle Spectrograph instrument \citep[EXES,][]{Richter2018} aboard SOFIA observes in MIR from 5 to 28 \micron\ with a spectral resolution of $10^3$ to $10^5$. Currently, this is the only spectrograph able to resolve individual molecular transitions over the entire MIR.  Several studies of hot cores with  SOFIA/EXES are ongoing \citep{Indriolo2015a,Barr2018,Dungee2018,Indriolo2020,Barr2020}. \citet{Rangwala2018} detected eight HCN R band rovibrational transitions towards IRc2 between 12.96 and 13.33 \micron, enough to directly calculate its temperature and column density. 

In this work, we present high-resolution observations of HCN, \hcniso, and HNC towards the hot core Orion IRc2 from 12.8 to 22.9 \micron. This is the first ISM detection of HNC and \hcniso\ in the MIR. Additionally, at MIR wavelengths, SOFIA’s beam size (3\farcs2$\times$3\farcs2) is much smaller compared with space-based missions like ISO (14\arcsec$\times$27\arcsec) and most previous sub-mm/mm studies from the ground.  This enables us to unambiguously isolate a previously unprobed, hotter component of molecular gas, traced by HCN, closest to the hot core from the surrounding emission.

 We provide our observational methods in \S \ref{sec:obs}, detail our analysis in \S \ref{sec:ana}, discuss the implications of the results and compare them with both other observations and models in \S \ref{ssec:dis}, and summarize our conclusions in \S \ref{sec:conc}. Our observations are part of a wider molecular survey of Orion IRc2 in the MIR from 7.2 to 8 and 13.2 to 28.3 \micron\ (publication in preparation).

\section{Observations and Data Reduction} \label{sec:obs}

\begin{figure*}
\begin{center}
\begin{tabular}{ccc}
\includegraphics[scale=0.39]{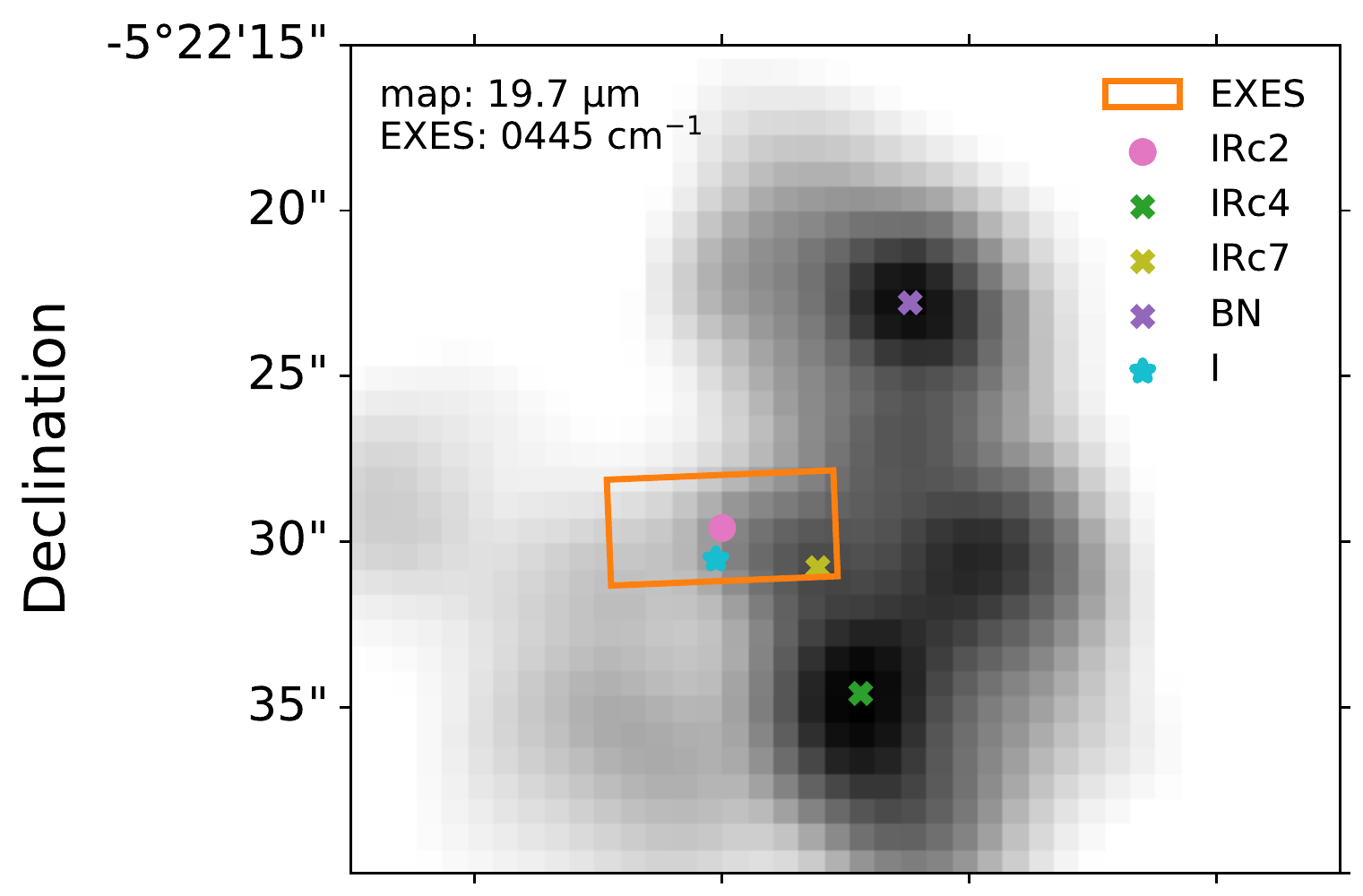}&\includegraphics[scale=0.39]{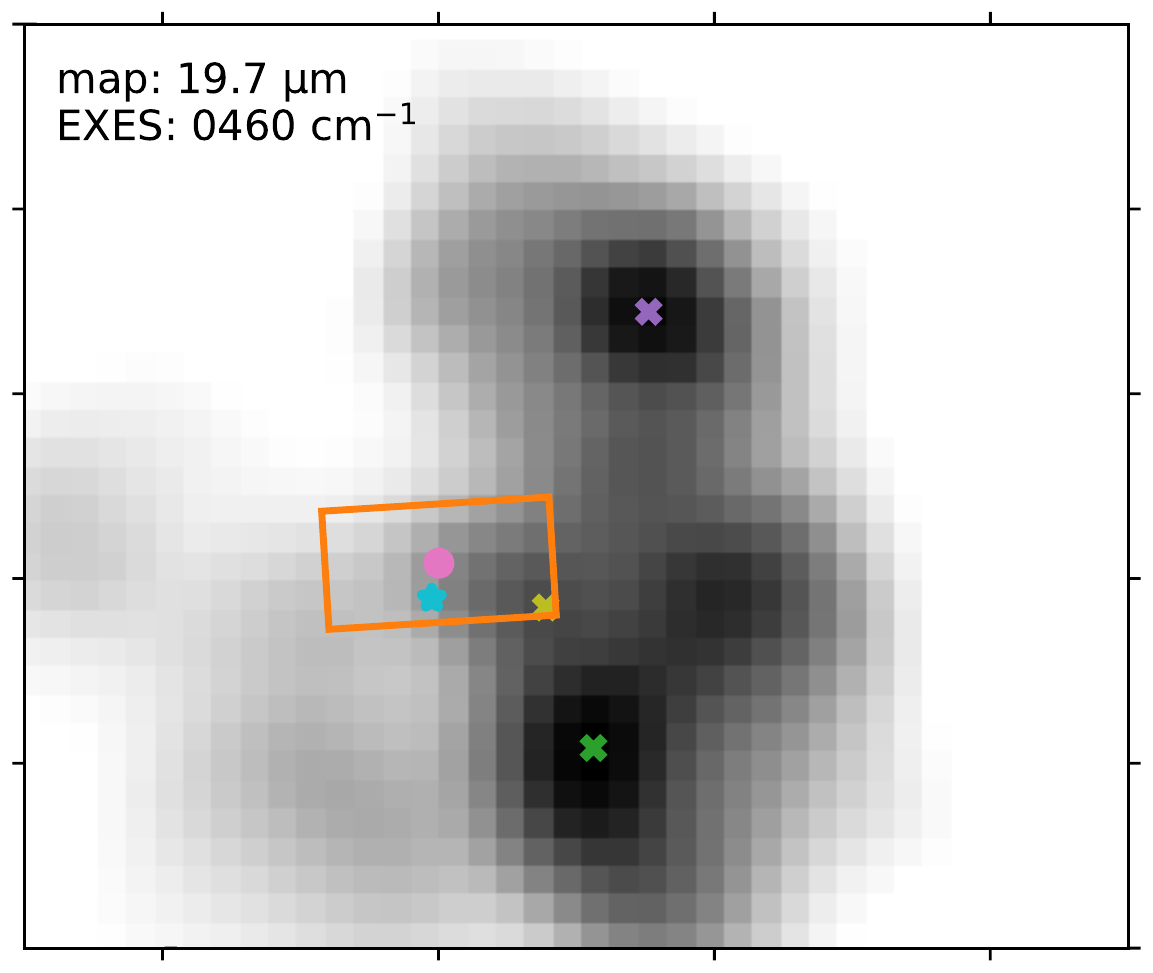}&\includegraphics[scale=0.39]{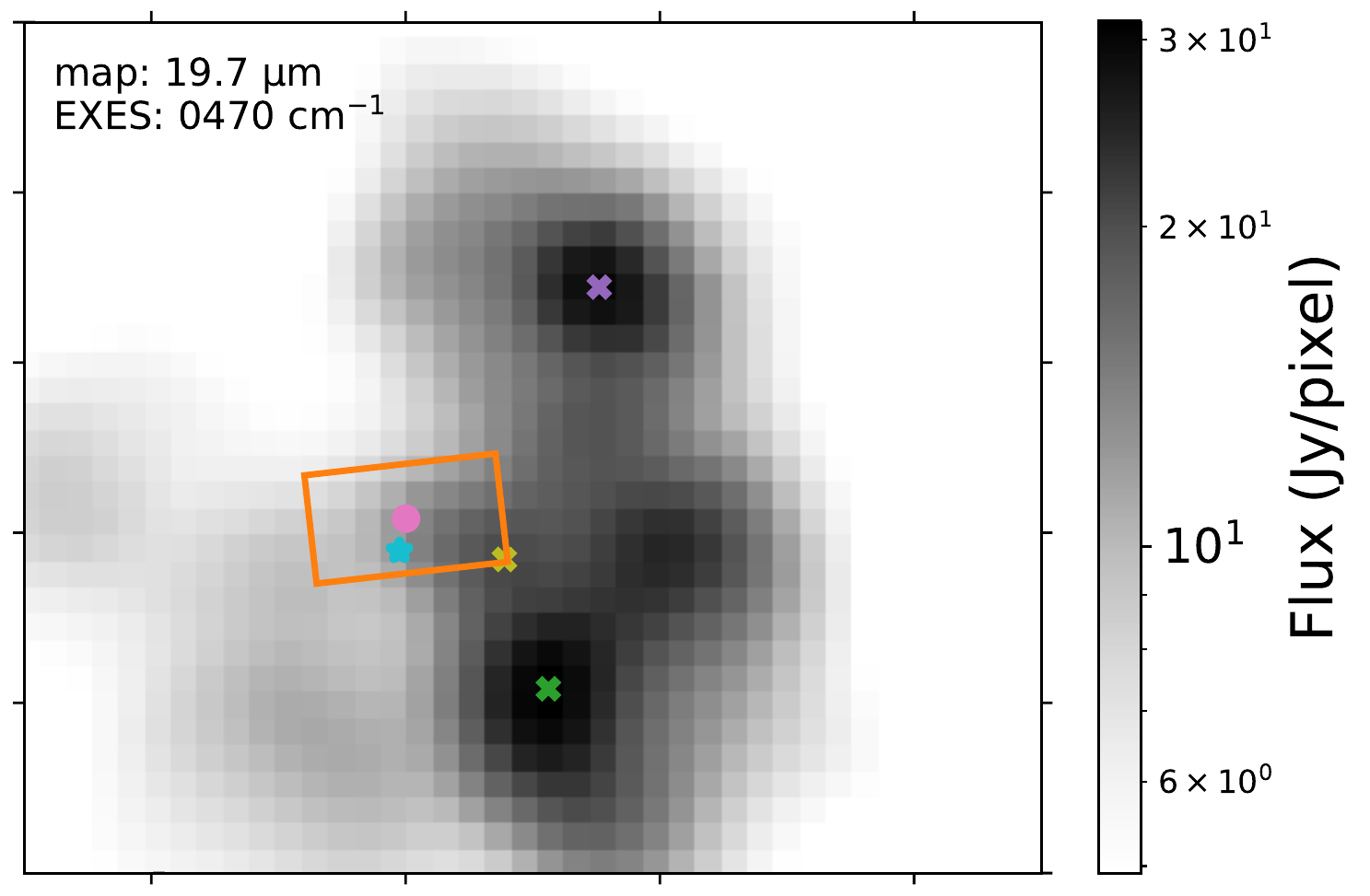}\\
\includegraphics[scale=0.39]{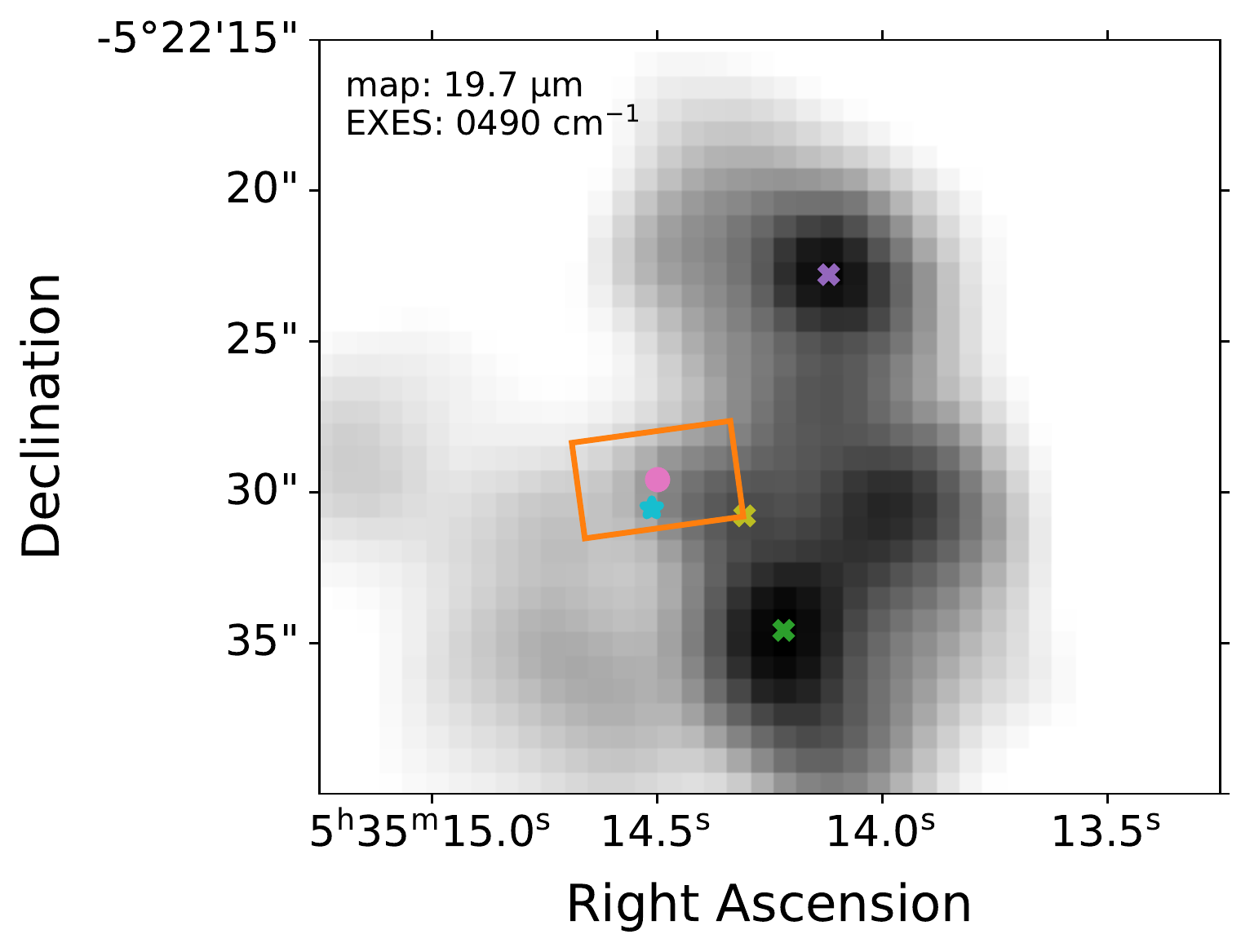}&\includegraphics[scale=0.39]{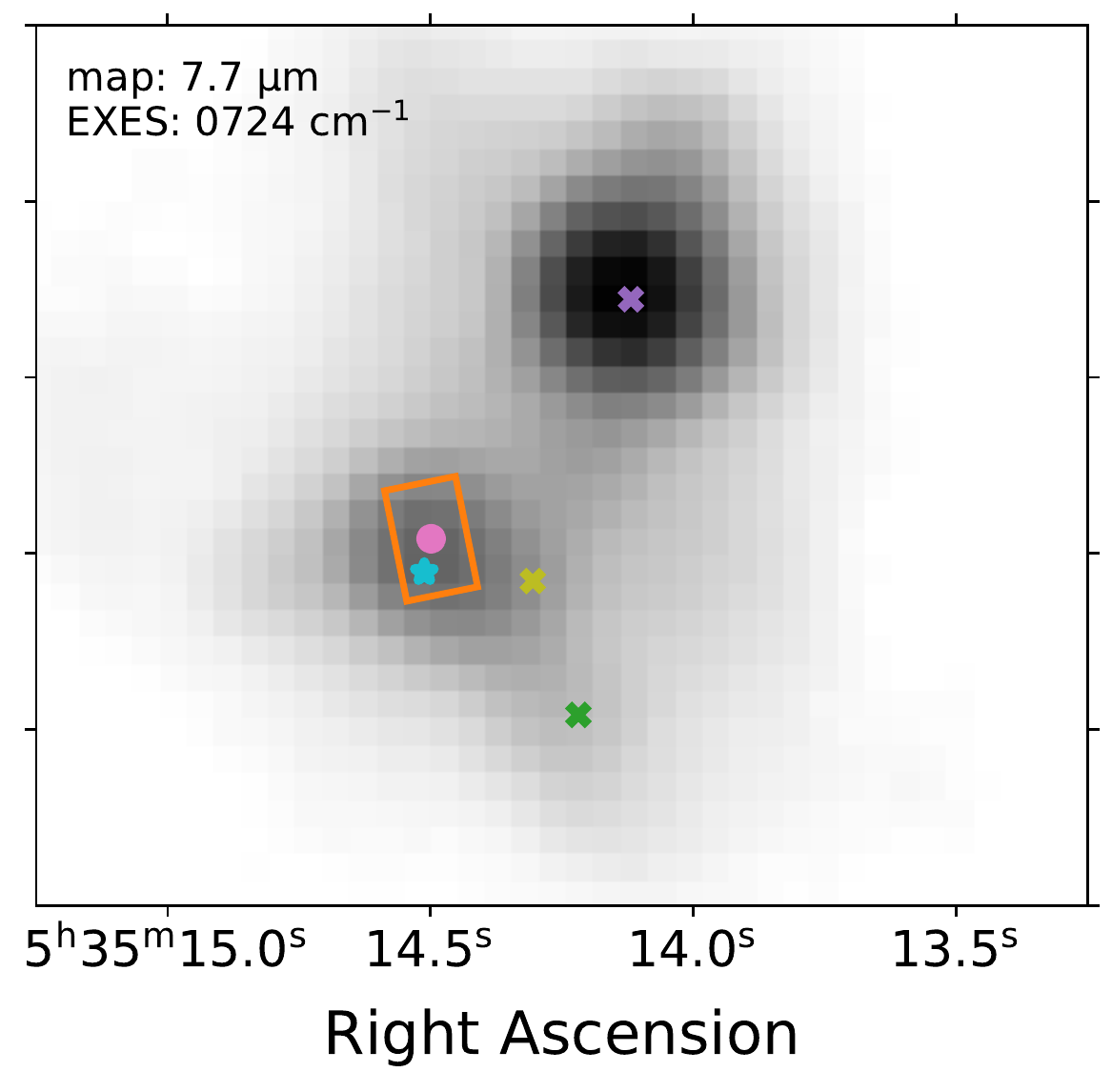}&\includegraphics[scale=0.39]{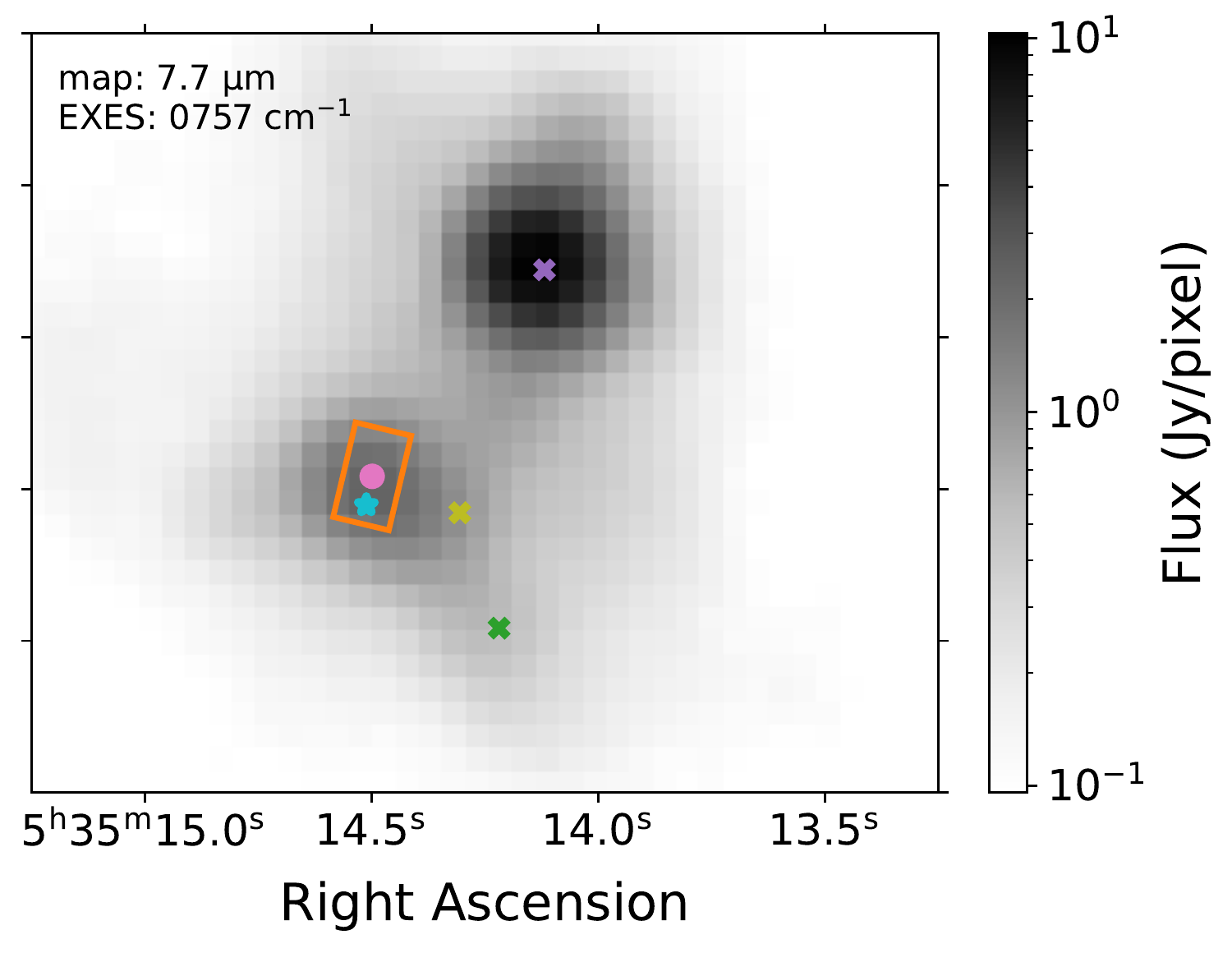}
\end{tabular}
\end{center}
\caption{The EXES footprint for each setting superimposed on an SOFIA/FORECAST map of the region \citep{DeBuizer2012}. In addition to our target, IRc2, the positions of radio source I and nearby IR sources IRc4, IRc7, and BN are given. Note in order to display maps closest in wavelength to each EXES setting, that 7.7 \micron\ map is for the 700 settings, and the 19.7 \micron\ map is for the 400's \ci\ setting. Their flux scales differ, given by the greyscale bars on the right.  \label{fig:footprint}}
\end{figure*}

We observed Orion IRc2 with the EXES instrument aboard the SOFIA observatory between 2018 October 27 and 31 at altitudes from about 42,000 to 44,000 ft in High-low mode. Spectra were acquired in the cross-dispersed high-resolution mode with a slit width of $3\farcs2$ giving a resolving power of about 60,000 ($\sim$ 5 \kms). We used the cross-disperser grating in 1st order to obtain the broadest simultaneous wavelength coverage per spectral setting. The length of the slit varied between $1\farcs9$ and $6\farcs9$, depending on the spectral setting. Table \ref{tab:obs} gives the details for these six settings, each of which is split into several orders. For all observations, we nodded the telescope to an off-source position relatively free of emission 15\arcsec\ East  and 25\farcs9 North of IRc2, at 1 minute intervals, in order to remove sky emission and thermal background from the telescope system.

\begin{deluxetable*}{cccccccc}
\tablecaption{Specifications for each EXES setting\label{tab:obs}}
\tablehead{\colhead{Setting} & \colhead{Species} & \colhead{Min $\lambda$} & \colhead{Max $\lambda$} &
\colhead{Date} & \colhead{Altitude} & \colhead{Slit Length$\times$Width} & \colhead{Integration Time} \\
\colhead{(\ci)} & \colhead{} & \colhead{(\micron)} & \colhead{(\micron)} &
\colhead{(yyyy-mm-dd)} & \colhead{(ft)} & \colhead{(\arcsec $\times$ \arcsec)} & \colhead{(s)}}
\startdata
757 & HCN & 12.8 & 13.6 & 2018-10-31 & 42,008 & $1.9\times 3.2$ &2,560\\
724 & HCN, \hcniso & 13.5 & 14.3 & 2018-10-30 & 44,002 & $2.1\times 3.2$& 2,880\\
490 & HNC & 20.1 & 20.8 & 2018-10-27 & 43,014 & $5.3\times 3.2$& 576\\
470 & HNC & 20.8 & 21.5 & 2018-10-27 & 43,015 & $5.7\times 3.2$& 512\\
460 & HNC & 21.5 & 22.2 & 2018-10-27 & 42,010 & $6.2\times 3.2$& 512\\
445 & HNC & 22.2 & 22.9 & 2018-10-27 & 42,005 & $6.9\times 3.2$& 512\\
\enddata
\end{deluxetable*}

We reduce the EXES data with the SOFIA Redux pipeline \citep{Clarke2015}. Wavelength scales are calibrated using sky emission line spectra produced for each setting by omitting the nod subtraction step and then adjusting the scale to match observed sky emission line wavelengths to their values in the HITRAN database \citep{Gordon2017}. The absolute velocity uncertainty is 0.15 to 0.3 \kms, estimated from High-Medium  mode sky lines of other settings. Figure \ref{fig:footprint} gives the EXES footprint for each observational setting over SOFIA/FORECAST maps at 7.7 and 19.7 \micron\ \citep{DeBuizer2012}. Each observation is centred over IRc2, and we also show the positions of IR sources IRc4, IRc7, and BN, and radio source I, which has  no IR component and may be heating IRc2. For settings 445 and 460 \ci, in particular, we were concerned about contamination and dilution from IRc7. However, after examining spectra split along the slit width, we confirm that the spectral lines originate from the slit centre, over IRc2, and not the edge over IRc7. Furthermore, from ground-based MIR spectroscopy of IRc2 and IRc7, \citet{Evans1991} found that \acet\ and HCN have 2 and 3 times higher column densities, respectively, towards IRc2 than IRc7. Therefore, our observations are centred over IRc2 and are unlikely to be contaminated by IRc7.

\begin{figure*}
\centering
\gridline{\fig{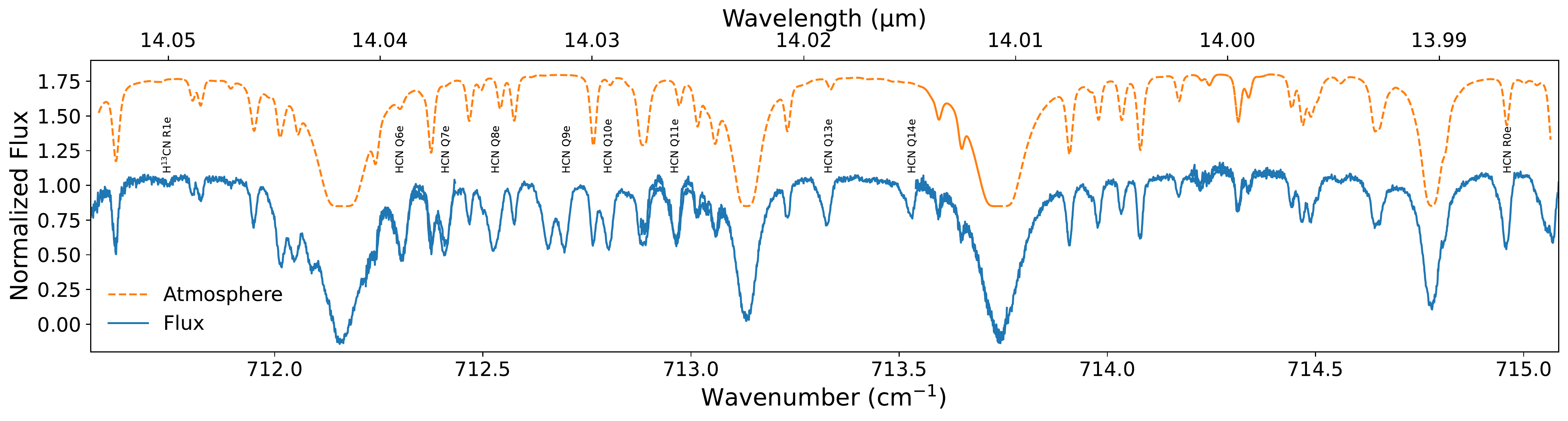}{0.95\textwidth}{}}
\vspace{-10mm}
\gridline{\fig{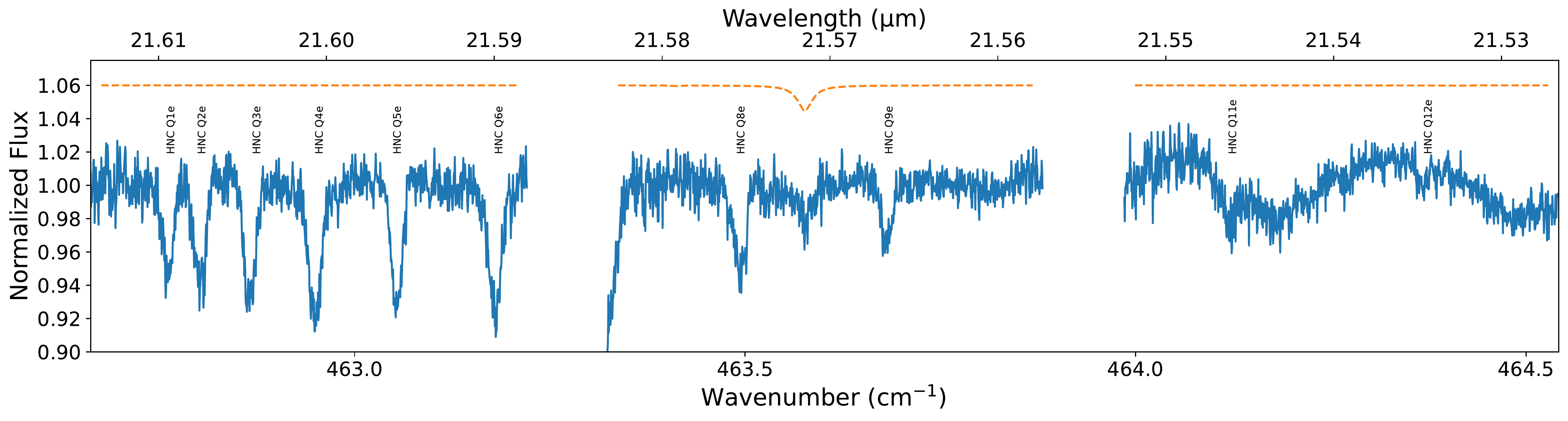}{0.95\textwidth}{}}
\vspace{-10mm}
\caption{Sample of EXES spectra for HCN (top), \hcniso\ (top), and HNC (bottom). The flux is normalized and the atmosphere smoothed according to \S \ref{ssec:atmos} and offset for display purposes. Note that in the top panel, lines overlap from adjacent orders. The bottom panel shows three separated orders, the leftmost two of which have been baseline corrected by a polynomial, while the rightmost has not due to the absence of a satisfactory polynomial.\label{fig:spec}}
\end{figure*}

We find HCN in two settings covering 701 to 783 \ci\ (12.8 to 14.3 \micron), \hcniso\ in one setting from 701 to 725 \ci\ (13.5 to 14.3 \micron), and HNC in four settings from 436 to 498 \ci\ (20.1 to 22.9 \micron). We searched for, but do not observe, the isotopologues HC$^{15}$N and DCN with the HITRAN \citep{Gordon2017} and GEISA \citep{Jacquinet-Husson2016} databases respectively and HN$^{13}$C and H$^{15}$NC with our own theoretically calculated line lists. Figure \ref{fig:spec} shows an example of spectra for the 724 and 460 \ci\ settings with lines of HCN, \hcniso, and HNC.

\section{Analysis} \label{sec:ana}

\subsection{Peak Finding} \label{ssec:peak}

In order to normalize our fluxes we must first identify the peaks and the baseline of each order. We employ a simple peak-finding algorithm to identify minima and maxima. For a given order, we perform Gaussian smoothing on the flux and then take the sign of the derivative of the smoothed flux. We consider blocks of consecutive pixels with the same sign. If two blocks of opposite signs are adjacent, or have a small gap between them, they are considered to belong to a peak and the peak position is taken to be the centre of the gap between them. If a block is next to an order endpoint, then that endpoint is considered a peak. Three input parameters to the peak-finder, on occasion, require adjustment depending on the order: sigma for strength of Gaussian smoothing, number of consecutive pixels with the same sign to be considered a block, and the size of the gap between blocks of consecutive pixels.

\subsection{Normalization and Atmospheric Line Correction} \label{ssec:atmos}

While the lines in settings 445, 460, 470, and 490 \ci\ are clear of atmospheric features, many lines in 724 and 757 \ci\ are mixed with atmospheric lines, necessitating correction (for example: HCN Q8e, top panel Figure \ref{fig:spec}). We download the unsmoothed ATRAN atmospheric model \citep{Lord1992}\footnote{\url{https://atran.arc.nasa.gov/cgi-bin/atran/atran.cgi}} for each setting given their altitude, latitude, and zenith of observation. This unsmoothed ATRAN model corresponds to infinite resolution. To divide the ATRAN from our flux, we seek to smooth  ATRAN to EXES's resolution and normalize our flux to ATRAN simultaneously. 

Each order requires a different normalization, especially the settings 445 to 490 \ci\, while an entire setting should have the same resolution. To achieve this, we implement the following procedure on the fluxes from each order:
\begin{enumerate}
    \item For each individual order, we run the peak finder described in \S \ref{ssec:peak} on the observed EXES flux and the ATRAN model.
    \item Due to standing waves in settings 445 and 460 \ci, a few orders require baseline correction where possible (see Figure \ref{fig:spec}, bottom panel). Considering all pixels not in a peak to be part of the baseline, we fit the baselines of these orders to a polynomial, $n_b$, and divide the flux by this polynomial in order to straighten the baseline.
    \item We mask peaks that appear in the EXES flux and not in the ATRAN model, as well as deep CO$_2$ lines that bottomed out at 0 in ATRAN but extended to negative in EXES (see Figure \ref{fig:spec}, top panel, for three examples). This leaves us with fluxes $F_{A}$ from ATRAN and $F_{E}$ from EXES that match.
    \item We conduct the following procedure on each order:
\begin{equation}
    \texttt{min}\Big[\sum_{i}^{\mathrm{pixels}} \Big\Vert \texttt{G1d}[F_{A,i},\sigma_G]- \frac{F_{E,i}}{n_{b,i}n_{\mathrm{c}}}\Big\Vert \Big],
    \label{eqn:opto}
\end{equation}
where $F_{A,i}$ is the ATRAN model flux at pixel $i$, $F_{E,i}$ is the our observed EXES flux at pixel $i$, $n_{b,i}$ is the normalization polynomial's value at pixel $i$ if needed (otherwise $n_{b,i}=1$, see step 2 above), $n_c$ is the normalization constant, \texttt{min} is the routine \texttt{optimize.minimize} from the \texttt{scipy} python package \citep{Virtanen2020} that minimizes its enclosed function, and \texttt{G1d} is the routine \texttt{scipy.ndimage.gaussian\_filter1d} that smooths its enclosed function with standard deviation $\sigma_G$. $F_{A,i}$, $F_{E,i}$, and $n_{b,i}$ are inputs to Equation \ref{eqn:opto}, and \texttt{min} returns the values of $n_c$ and $\sigma_G$ that minimize Equation \ref{eqn:opto}. We are finding the resolution and flux normalization that minimizes the difference between the ATRAN model and our observed EXES flux. 
\item We retain $n_c$ for each individual order, but discard $\sigma_G$, because many orders do not have enough, or any, ATRAN lines to match accurately the resolution. With each order normalized individually, we input the fluxes from an entire setting back into Equation \ref{eqn:opto} to obtain that setting's overall $\sigma_G$. 

\item We divide the order-normalized EXES flux by the setting-wide smoothed ATRAN model to correct for the atmosphere and use this flux for line fitting in \S \ref{ssec:line}.
\end{enumerate}

\vspace{2mm}
\subsection{Line Fitting} \label{ssec:line}

We rerun the peak finder from \S \ref{ssec:peak} again on the normalized flux to match these peaks with the molecular absorption lines of interest. In many cases, this is sufficient to correctly identify the lines' extent. We only need to manually identify line boundaries in cases where the line was either weak or one of many in a crowded region. For settings 724 and 757 \ci, we also correct for atmosphere as in \S \ref{ssec:atmos}, while this is unnecessary for the other settings as no lines are near to any atmospheric absorption. 

With \texttt{scipy.optimize.curvefit}, we fit the measured absorption lines with a Gaussian profile following \citet{Indriolo2015}:
\begin{equation}
    I = I_0 e^{-\tau_0 G},
\end{equation}
where,
\begin{equation}
    G = \mathrm{exp}\Big[-\frac{(v-v_c)^2}{2\sigma^2_v}\Big],    
\end{equation}
$I_0$ is the normalized continuum level (typically close to unity), $\tau_0$ is the line centre optical depth, $v$ is the local standard of rest (LSR) velocity, $v_c$ is the velocity of the line centre, and $\sigma_v$ is the velocity dispersion. In the cases of double Gaussians, 1 and 2, we fit: 
\begin{equation}
    I = I_0 e^{-(\tau_{01} G_1 + \tau_{02} G_2)}.   
\end{equation}
Triple Gaussians follow a similar equation. Each HNC and \hcniso\ line is best fit by a single Gaussian. Almost all HCN lines show two velocity components and are best fit by a double Gaussian with a handful of exceptions. P3e, R17e, and R18e are not resolved enough for a double Gaussian. Q9e and R12e overlap with acetylene absorption lines, leaving single Gaussians. Q8e, Q11e, Q14e and R6e overlap with atmospheric lines that are divided out. However, due to the ATRAN model not matching the data exactly, the remaining flux is slightly distorted and a triple Gaussian fits these lines best to correct for this distortion.

\begin{figure*}
\centering
\gridline{\fig{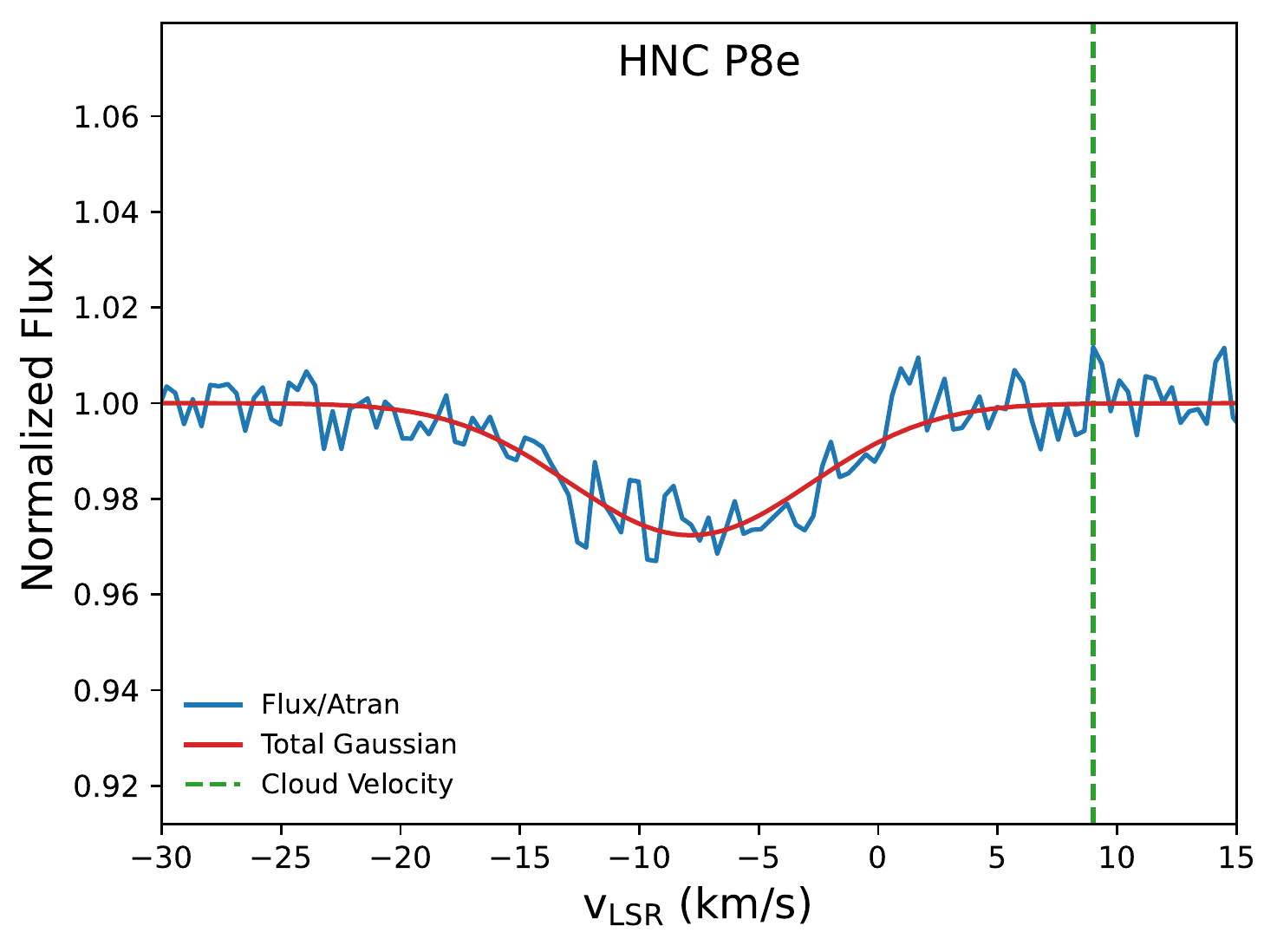}{0.32\textwidth}{}\hspace{-5mm}
          \fig{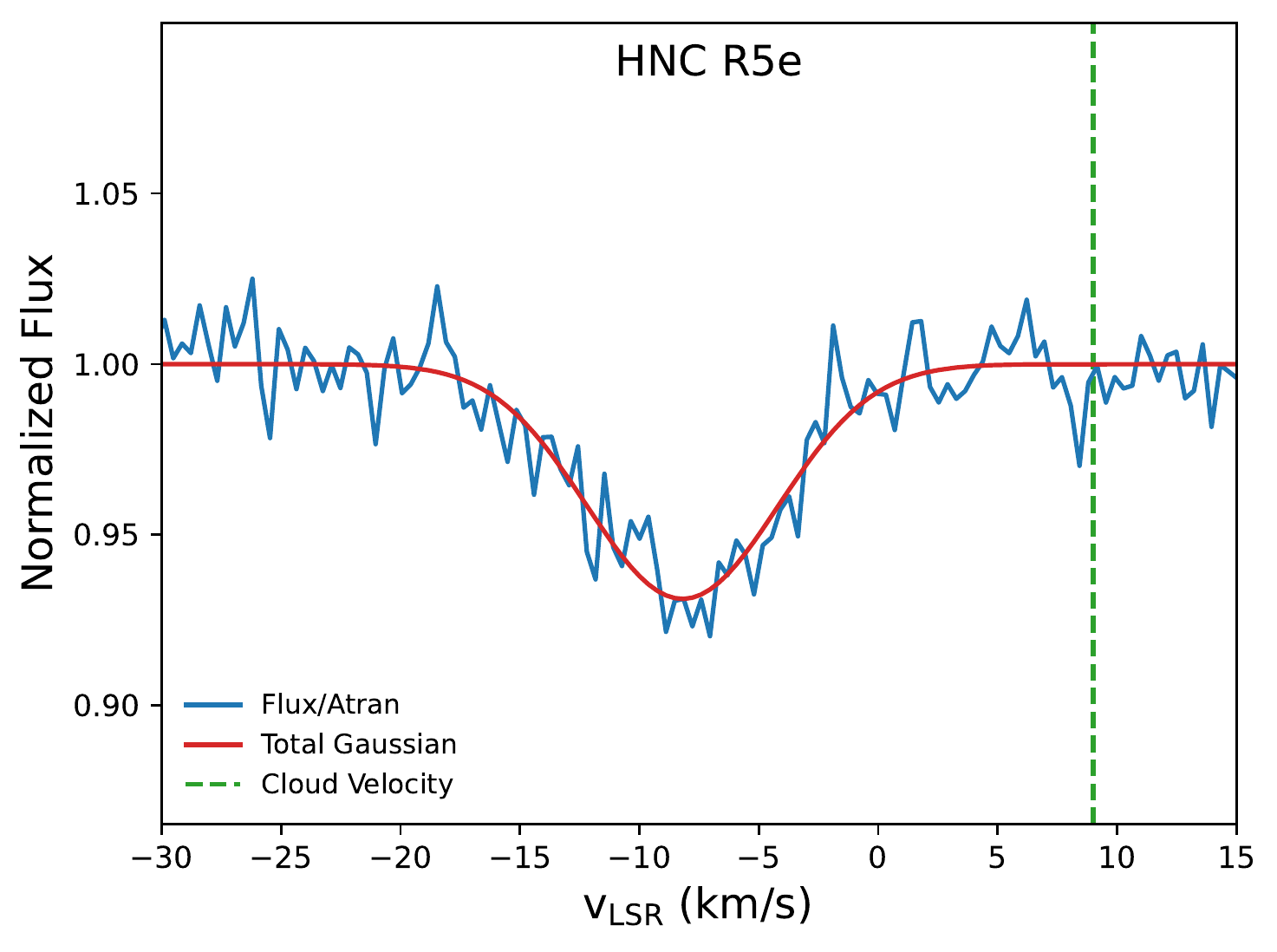}{0.32\textwidth}{}\hspace{-5mm}
          \fig{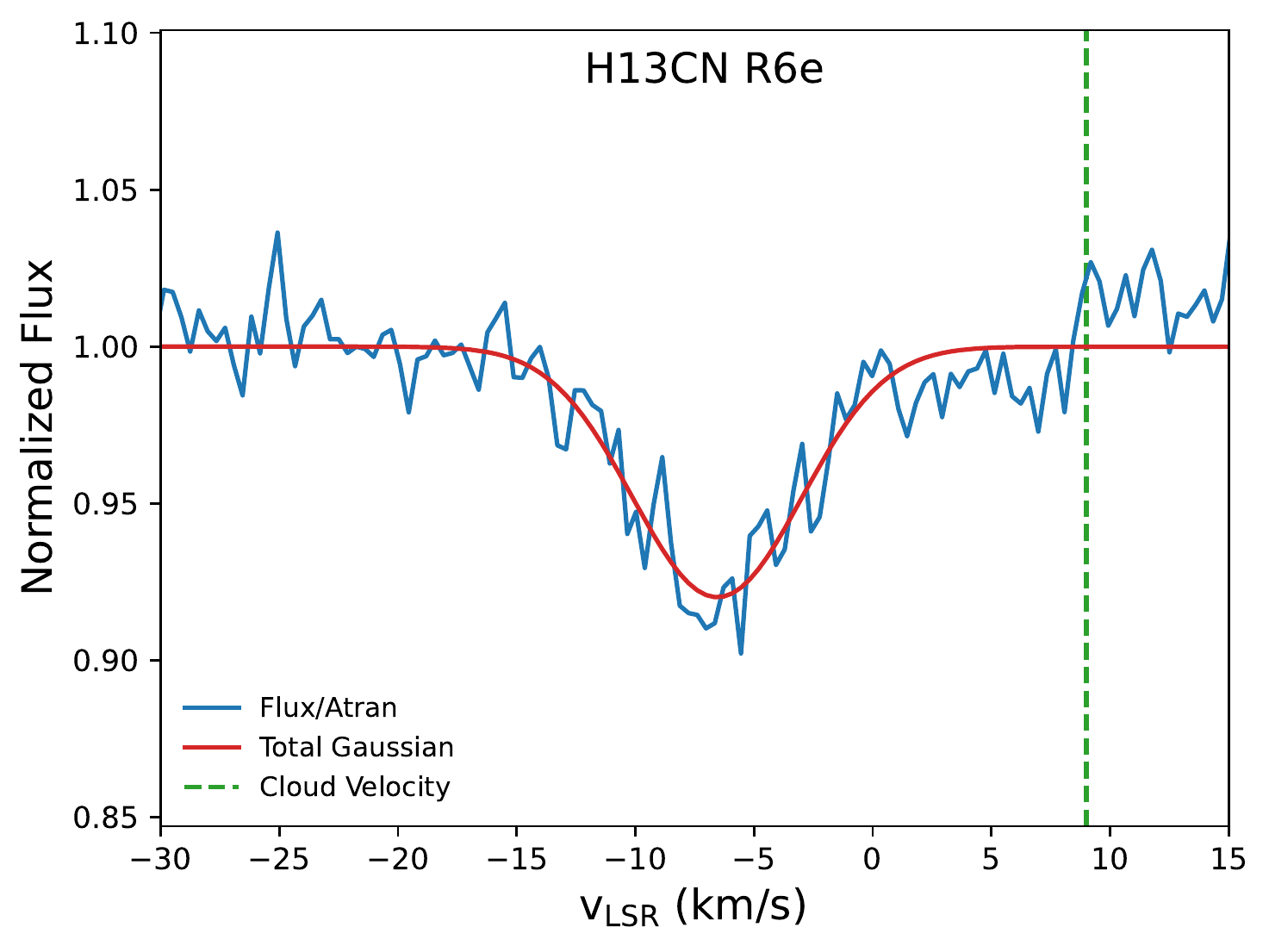}{0.32\textwidth}{}}
\vspace{-10mm}
\gridline{\fig{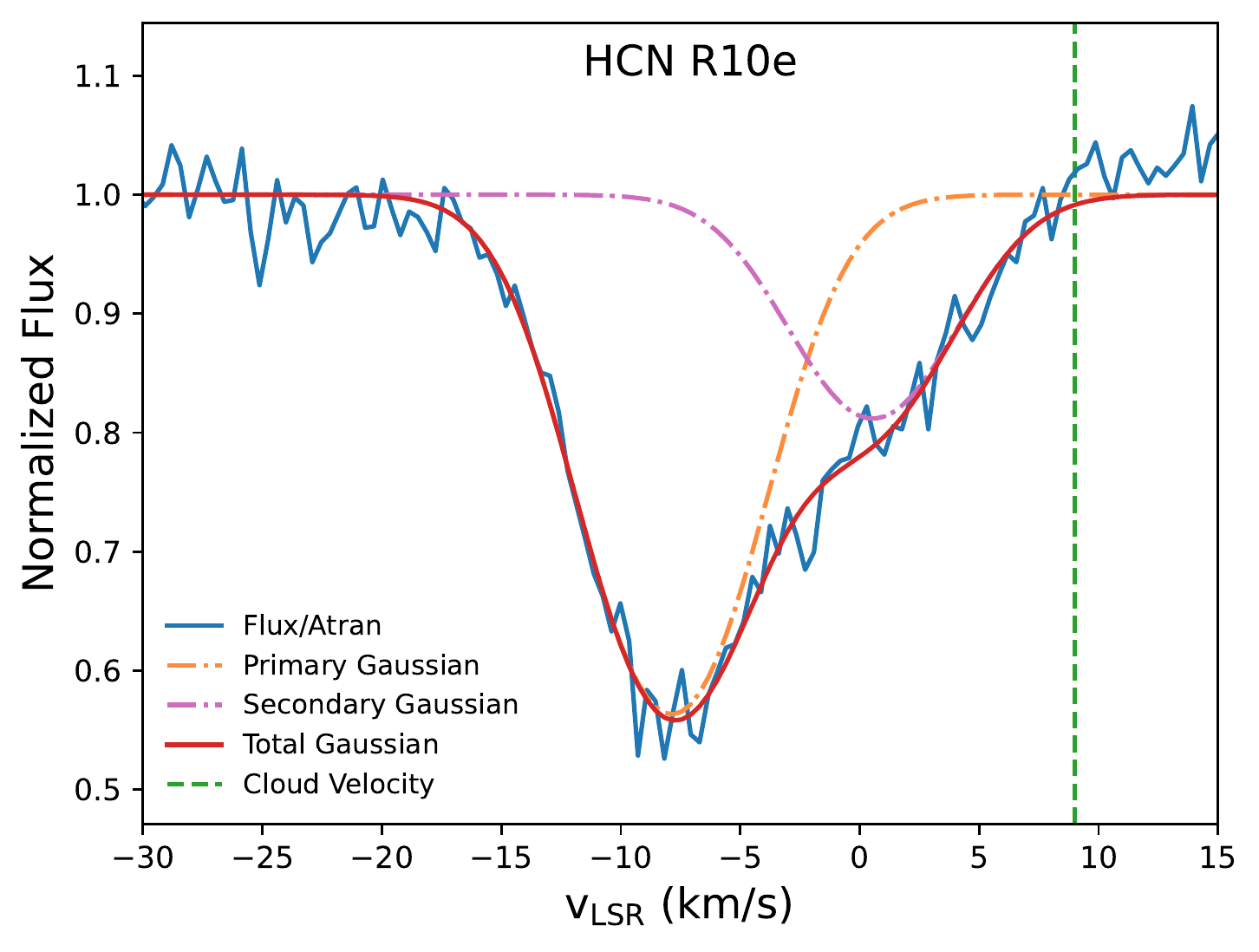}{0.32\textwidth}{}\hspace{-5mm}
          \fig{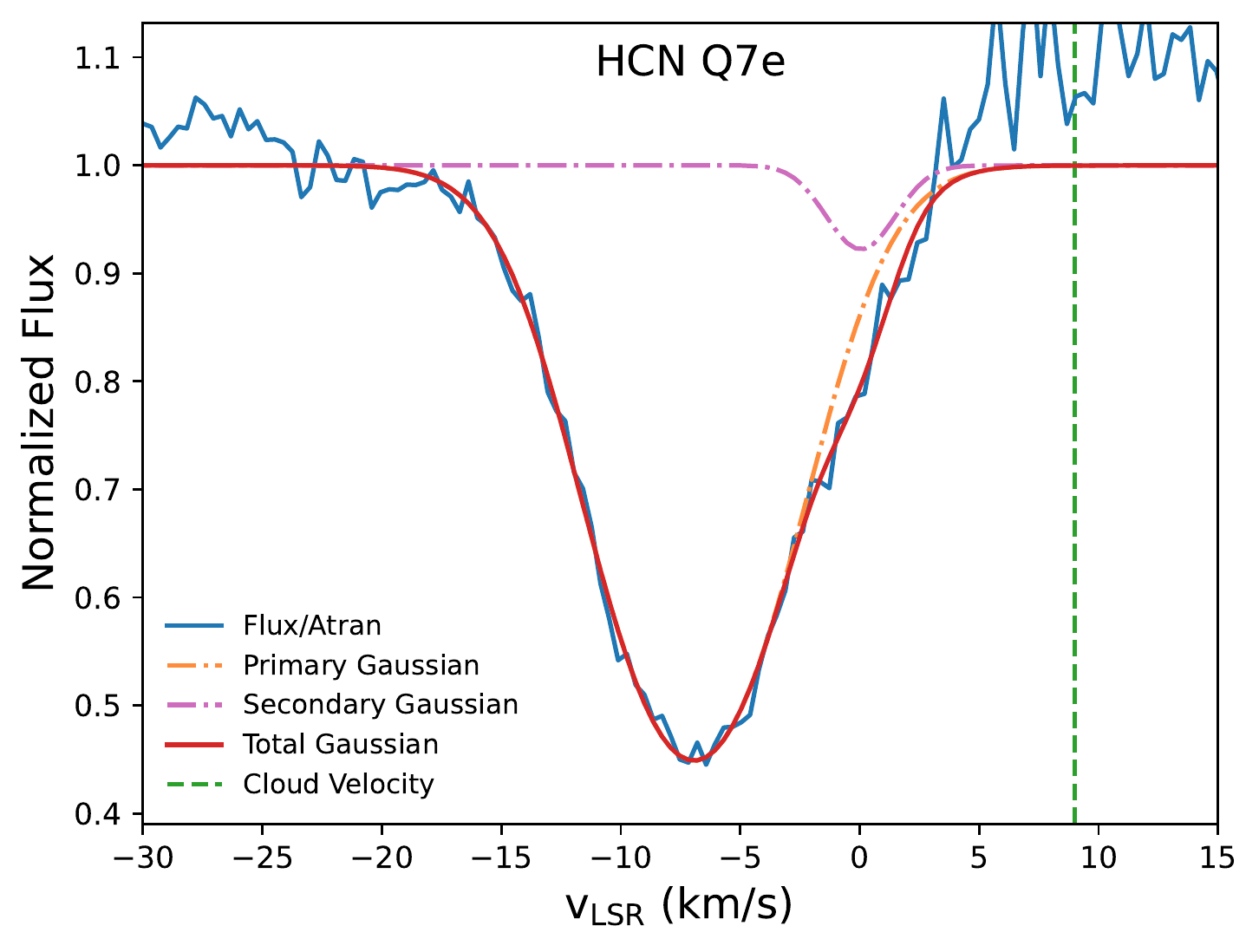}{0.32\textwidth}{}\hspace{-5mm}
          \fig{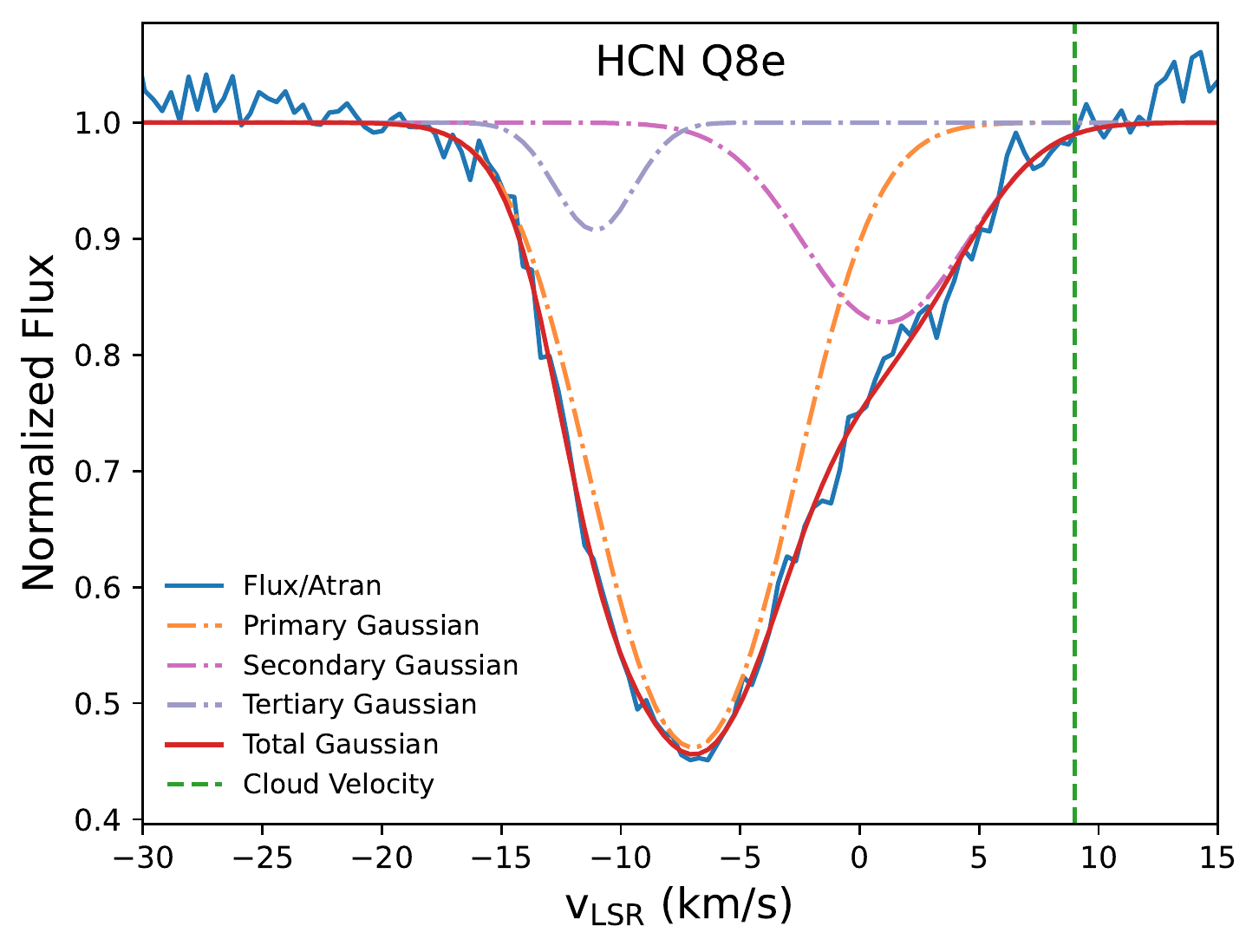}{0.32\textwidth}{}}
\vspace{-10mm}
\caption{Gaussian fits for selected HNC (top left two), \hcniso\ (top right), and HCN (bottom row) lines normalized to their baselines. The HCN and \hcniso\ fluxes have been corrected for atmospheric absorption. HNC and \hcniso\ are best fit by a single Gaussian, while HCN is best fit by a double Gaussian showing two velocity components. For the case of Q8e, two Gaussians are needed for the main velocity component to catch distortion from atmospheric division. The main velocity component of HCN is similar to that of HNC and \hcniso. The vertical dotted line indicates the systematic, ambient cloud velocity 9 \kms\  \citep{Zapata2012} \label{fig:gauss}}
\end{figure*}

Figure \ref{fig:gauss} shows six examples of fits, with single Gaussians for HNC and \hcniso, and double or triple Gaussians for HCN. HCN R10e is an example of where the secondary velocity component is distinct. While HCN Q7e is similarly best fit by a double Gaussian, an atmospheric line falls over the secondary component and little remains after atmospheric correction. In cases such as this, we consider the entire line as belonging to the primary velocity component, even though some trace of the secondary remains. Q8e is an example of a triple Gaussian, where even after atmospheric correction the primary Gaussian is distorted and two Gaussians fit it best. Another Gaussian fits the secondary velocity component.

We calculate the column density, $N_l$, in the lower state of an observed transition from the integral over the absorption line:
\begin{eqnarray}
    dN_l/dv &=& \frac{g_l}{g_u}\frac{8\pi}{A\lambda^3}\tau_0 G,\nonumber \\
    N_l &=& \sqrt{2\pi} \frac{g_l}{g_u}\frac{8\pi}{A\lambda^3} \tau_0 \sigma_v,
\end{eqnarray}
where $g_l$ and $g_u$ are the lower and upper statistical weights respectively, $A$ is the Einstein constant for spontaneous emission, and $\lambda$ is the rest wavelength of the transition. We obtain these parameter values (along with $E_l$ required in \S \ref{ssec:rot}) for HCN and \hcniso\ from the HITRAN database \citep{Gordon2017} and for HNC from the GEISA database \citep{Jacquinet-Husson2016}.

\begin{deluxetable*}{rrrrrrrrr}
\tablecaption{Observed $\nu_2$ band HNC Transitions and Inferred Parameters\label{tab:HNClines}}
\tablehead{
\colhead{Transition} & \colhead{Wavenumber} & \colhead{$E_l/k_b$} & \colhead{$g_l$} & \colhead{$A$} & \colhead{\vlsr} & \colhead{\vfwhm} & \colhead{$\tau_0$} & \colhead{$N_l$} \\
\colhead{} & \colhead{(\ci)} & \colhead{(K)} & \colhead{} & \colhead{(\si)} & \colhead{(\kms)} & \colhead{(\kms)} & \colhead{} & \colhead{$\times10^{14}$\csi}}
\startdata
P3e&453.64798&26.1&42&1.269&--6.8$\pm$0.3&12.6$\pm$1.5&0.031$\pm$0.003&1.09$\pm$0.20\\
P5e&447.5964&65.3&66&1.352&--7.0$\pm$0.2&10.6$\pm$0.5&0.041$\pm$0.001&0.94$\pm$0.05\\
P6e&444.57022&91.4&78&1.353&--5.3$\pm$0.4&14.4$\pm$2.0&0.029$\pm$0.003&0.86$\pm$0.19\\
P7e&441.54388&121.8&90&1.345&--9.1$\pm$0.3&10.5$\pm$1.3&0.031$\pm$0.003&0.65$\pm$0.11\\
P8e&438.51747&156.6&102&1.33&--7.9$\pm$0.3&11.8$\pm$1.1&0.028$\pm$0.002&0.64$\pm$0.08\\
Q1e&462.74319&4.4&18&3.379&--8.9$\pm$0.2&10.3$\pm$0.7&0.057$\pm$0.003&0.46$\pm$0.04\\
Q2e&462.78519&13.1&30&3.379&--8.1$\pm$0.2&12.8$\pm$1.2&0.071$\pm$0.004&0.71$\pm$0.10\\
Q3e&462.84818&26.1&42&3.38&--8.9$\pm$0.1&11.7$\pm$0.6&0.080$\pm$0.003&0.73$\pm$0.06\\
Q4e&462.93214&43.5&54&3.382&--8.4$\pm$0.2&11.6$\pm$0.6&0.080$\pm$0.003&0.72$\pm$0.05\\
Q5e&463.03705&65.3&66&3.383&--8.3$\pm$0.1&10.7$\pm$0.4&0.082$\pm$0.002&0.69$\pm$0.04\\
Q6e&463.16287&91.4&78&3.385&--8.4$\pm$0.2&13.2$\pm$0.5&0.078$\pm$0.003&0.81$\pm$0.04\\
Q8e&463.47714&156.6&102&3.39&--7.7$\pm$0.2&12.6$\pm$1.0&0.055$\pm$0.003&0.54$\pm$0.06\\
Q9e&463.66551&195.8&114&3.393&--7.7$\pm$0.3&10.5$\pm$1.0&0.037$\pm$0.002&0.30$\pm$0.04\\
Q11e&464.10446&287.1&138&3.4&--8.1$\pm$0.5&7.7$\pm$2.5&0.022$\pm$0.005&0.13$\pm$0.07\\
Q12e&464.35494&339.3&150&3.404&--5.9$\pm$0.6&8.6$\pm$1.6&0.013$\pm$0.002&0.09$\pm$0.02\\
R0e&465.74576&0.0&6&2.298&--8.3$\pm$0.3&11.9$\pm$1.3&0.037$\pm$0.003&0.17$\pm$0.03\\
R1e&468.76863&4.4&18&2.113&--8.0$\pm$0.2&14.6$\pm$0.7&0.069$\pm$0.003&0.79$\pm$0.05\\
R2e&471.7907&13.1&30&2.053&--8.1$\pm$0.2&12.3$\pm$0.8&0.076$\pm$0.003&0.92$\pm$0.08\\
R3e&474.8119&26.1&42&2.037&--8.0$\pm$0.2&11.3$\pm$0.6&0.081$\pm$0.003&1.01$\pm$0.08\\
R5e&480.85136&65.3&66&2.055&--8.2$\pm$0.2&9.3$\pm$0.5&0.071$\pm$0.003&0.81$\pm$0.04\\
R7e&486.88636&121.8&90&2.102&--7.7$\pm$0.3&12.4$\pm$1.0&0.048$\pm$0.003&0.78$\pm$0.07\\
R8e&489.902&156.6&102&2.131&--6.2$\pm$0.3&9.2$\pm$1.0&0.042$\pm$0.003&0.51$\pm$0.08\\
R9e&492.91629&195.8&114&2.163&--7.8$\pm$0.3&12.9$\pm$0.9&0.043$\pm$0.002&0.75$\pm$0.06\\
R10e&495.92916&239.3&126&2.197&--5.9$\pm$0.5&8.4$\pm$1.3&0.026$\pm$0.004&0.30$\pm$0.07\\
\enddata
\tablecomments{Wavenumber is the rest wavenumber of the transition, $E_l$ is the energy level of the lower state, $k_b$ is the Boltzmann constant, $g_l$ is the lower statistical weight, $A$ is the Einstein constant, \vlsr\ is the observed local standard of rest velocity, \vfwhm\ is the observed full-width half-maximum, $\tau_0$ is the observed optical depth, and $N_l$ is the observed column density of the transition. Data in the first five columns are from the GEISA database \citep{Jacquinet-Husson2016}.}
\end{deluxetable*}

\begin{deluxetable*}{rrrrrrrrr}
\tablecaption{Observed $\nu_2$ band \hcniso\ Transitions and Inferred Parameters\label{tab:H13CNlines}}
\tablehead{
\colhead{Transition} & \colhead{Wavenumber} & \colhead{$E_l/k_b$} & \colhead{$g_l$} & \colhead{$A$} & \colhead{\vlsr} & \colhead{\vfwhm} & \colhead{$\tau_0$} & \colhead{$N_l$} \\
\colhead{} & \colhead{(\ci)} & \colhead{(K)} & \colhead{} & \colhead{(\si)} & \colhead{(\kms)} & \colhead{(\kms)} & \colhead{} & \colhead{$\times10^{14}$\csi}}
\startdata
R1e&711.72312&4.1&36&1.206&--7.5$\pm$0.4&8.7$\pm$0.9&0.059$\pm$0.005&2.46$\pm$0.27\\
R6e&726.098321&87.0&156&1.143&--6.5$\pm$0.2&8.2$\pm$0.6&0.083$\pm$0.005&5.30$\pm$0.44\\
R8e&731.839651&149.2&204&1.157&--5.7$\pm$0.3&7.0$\pm$0.8&0.054$\pm$0.005&3.05$\pm$0.40\\
\enddata
\tablecomments{See Table \ref{tab:HNClines} for column descriptions. Data in the first five columns are from the HITRAN database \citep{Gordon2017}.}
\end{deluxetable*}

Tables \ref{tab:HNClines}, \ref{tab:H13CNlines}, and \ref{tab:HCNlines} list the best-fit parameters with fitted errors as well as the relevant molecular database parameters for HNC, \hcniso, and HCN respectively. All three species transition from the ground state to the $\nu_2$ band. A number of lines are missing due to gaps between orders or heavy interference from atmospheric and acetylene lines. 

In the case of HCN, a superscript on the column density indicates to which velocity component the Gaussian fit belongs. These components separately contribute to rotation diagrams in \S \ref{ssec:rot}. For components with more than one Guassian fit, the column density is the total of the Gaussian fits comprising it, while the velocity will be the velocity of the component with the highest optical depth. This is because the shallower components are artifacts of atmospheric lines.

\startlongtable
\begin{deluxetable*}{rrrrrrrrr}
\tablecaption{Observed $\nu_2$ band HCN Transitions and Inferred Parameters\label{tab:HCNlines}}
\tablehead{
\colhead{Transition} & \colhead{Wavenumber} & \colhead{$E_l/k_b$} & \colhead{$g_l$} & \colhead{$A$} & \colhead{\vlsr} & \colhead{\vfwhm} & \colhead{$\tau_0$} & \colhead{$N_l$} \\
\colhead{} & \colhead{(\ci)} & \colhead{(K)} & \colhead{} & \colhead{(\si)} & \colhead{(\kms)} & \colhead{(\kms)} & \colhead{} & \colhead{$\times10^{14}$\csi}}
\startdata
P2e&706.0664&12.8&30&0.6576&--7.5$\pm$0.2&7.8$\pm$0.3&0.352$\pm$0.008&65.46$\pm$3.32$^1$\\
&&&&&0.7$\pm$0.3&7.5$\pm$0.6&0.173$\pm$0.009&30.99$\pm$3.46$^2$\\
P3e&703.109429&25.5&42&0.778&--6.8$\pm$0.3&11.1$\pm$1.0&0.429$\pm$0.029&79.92$\pm$8.57$^1$\\
Q6e&712.286&89.3&78&2.028&--7.0$\pm$0.1&9.7$\pm$0.3&0.750$\pm$0.017&34.76$\pm$1.11$^1$\\
&&&&&---&\textit{4.4$\pm$0.6}&\textit{0.180$\pm$0.019}&\textit{3.75$\pm$0.63$^1$}\\
Q7e&712.388056&119.1&90&2.028&--6.9$\pm$0.1&8.9$\pm$0.2&0.801$\pm$0.011&34.08$\pm$0.71$^1$\\
&&&&&---&\textit{3.4$\pm$0.8}&\textit{0.081$\pm$0.017}&\textit{1.29$\pm$0.43$^1$}\\
Q8e&712.504639&153.1&102&2.028&--7.0$\pm$0.3&8.3$\pm$0.3&0.772$\pm$0.018&30.55$\pm$1.60$^1$\\
&&&&&1.1$\pm$0.5&7.7$\pm$0.8&0.189$\pm$0.017&6.97$\pm$1.22$^2$\\
&&&&&---&\textit{3.8$\pm$1.1}&\textit{0.097$\pm$0.041}&\textit{1.76$\pm$1.18$^1$}\\
Q9e&712.635726&191.4&114&2.028&--7.4$\pm$0.1&12.4$\pm$0.2&0.692$\pm$0.010&41.06$\pm$0.68$^1$\\
Q10e&712.781294&233.9&126&2.027&--7.0$\pm$0.1&8.5$\pm$0.2&0.673$\pm$0.007&27.23$\pm$0.53$^1$\\
&&&&&---&\textit{4.6$\pm$0.6}&\textit{0.117$\pm$0.009}&\textit{2.58$\pm$0.37$^1$}\\
Q11e&712.941315&280.7&138&2.027&--7.4$\pm$0.1&8.1$\pm$0.3&0.592$\pm$0.011&23.04$\pm$0.75$^1$\\
&&&&&2.7$\pm$0.7&5.9$\pm$1.8&0.140$\pm$0.011&3.96$\pm$1.23$^2$\\
&&&&&---&\textit{3.2$\pm$1.0}&\textit{0.117$\pm$0.049}&\textit{1.78$\pm$1.21$^2$}\\
Q13e&713.304602&386.9&162&2.026&--7.3$\pm$0.2&7.2$\pm$0.3&0.419$\pm$0.015&14.52$\pm$0.99$^1$\\
&&&&&--0.1$\pm$0.7&7.6$\pm$1.0&0.134$\pm$0.013&4.87$\pm$1.00$^2$\\
Q14e&713.5078&446.5&174&2.026&--7.8$\pm$0.2&6.1$\pm$0.3&0.309$\pm$0.010&9.08$\pm$0.65$^1$\\
&&&&&0.2$\pm$0.7&8.0$\pm$1.2&0.109$\pm$0.007&4.18$\pm$0.81$^2$\\
&&&&&---&\textit{3.2$\pm$0.8}&\textit{0.086$\pm$0.032}&\textit{1.33$\pm$0.79$^2$}\\
R0e&714.935627&0.0&6&1.371&--6.7$\pm$0.3&9.7$\pm$0.5&0.438$\pm$0.011&10.12$\pm$0.63$^1$\\
&&&&&2.4$\pm$0.8&8.0$\pm$1.4&0.112$\pm$0.014&2.13$\pm$0.57$^2$\\
R1e&717.89124&4.3&18&1.251&--7.2$\pm$0.1&8.5$\pm$0.1&0.700$\pm$0.007&28.23$\pm$0.54$^1$\\
&&&&&1.4$\pm$0.2&6.5$\pm$0.4&0.152$\pm$0.007&4.66$\pm$0.40$^2$\\
R3e&723.800848&25.5&42&1.19&--7.2$\pm$0.1&9.4$\pm$0.3&0.779$\pm$0.019&48.37$\pm$2.25$^1$\\
&&&&&2.1$\pm$0.3&7.6$\pm$0.7&0.212$\pm$0.018&10.63$\pm$1.85$^2$\\
R4e&726.7547&42.5&54&1.184&--7.3$\pm$0.1&8.7$\pm$0.2&0.780$\pm$0.013&47.97$\pm$1.81$^1$\\
&&&&&1.1$\pm$0.6&7.5$\pm$1.1&0.154$\pm$0.014&8.20$\pm$1.69$^2$\\
R5e&729.70782&63.8&66&1.183&--7.3$\pm$0.1&11.6$\pm$0.2&0.755$\pm$0.013&65.02$\pm$1.25$^1$\\
R6e&732.660136&89.3&78&1.187&--7.2$\pm$0.2&6.7$\pm$0.8&0.698$\pm$0.091&35.76$\pm$8.44$^1$\\
&&&&&--1.0$\pm$1.4&11.0$\pm$1.6&0.272$\pm$0.042&22.99$\pm$6.66$^2$\\
&&&&&---&\textit{3.5$\pm$0.7}&\textit{0.183$\pm$0.073}&\textit{4.97$\pm$2.84$^1$}\\
R7e&735.611573&119.1&90&1.194&--7.5$\pm$0.2&8.3$\pm$0.2&0.760$\pm$0.018&49.65$\pm$2.16$^1$\\
&&&&&0.9$\pm$0.7&8.9$\pm$0.9&0.183$\pm$0.014&12.83$\pm$2.13$^2$\\
R10e&744.459871&233.9&126&1.221&--7.9$\pm$0.2&8.1$\pm$0.4&0.574$\pm$0.017&38.59$\pm$2.37$^1$\\
&&&&&0.5$\pm$0.6&7.9$\pm$0.9&0.208$\pm$0.017&13.56$\pm$2.32$^2$\\
R11e&747.407049&280.7&138&1.231&--7.3$\pm$0.3&9.9$\pm$0.5&0.468$\pm$0.013&38.86$\pm$2.47$^1$\\
&&&&&1.4$\pm$0.7&7.9$\pm$1.1&0.138$\pm$0.021&9.12$\pm$2.33$^2$\\
R12e&750.352972&331.7&150&1.242&--7.5$\pm$0.1&7.9$\pm$0.3&0.420$\pm$0.015&28.00$\pm$1.26$^1$\\
R13e&753.297564&386.9&162&1.254&--7.9$\pm$0.3&6.0$\pm$0.7&0.247$\pm$0.046&12.54$\pm$3.54$^1$\\
&&&&&--0.5$\pm$1.6&12.0$\pm$3.0&0.139$\pm$0.015&14.14$\pm$4.73$^2$\\
R15e&759.182446&510.2&186&1.278&--7.3$\pm$0.6&7.6$\pm$1.0&0.143$\pm$0.011&9.31$\pm$1.58$^1$\\
&&&&&1.3$\pm$0.8&8.4$\pm$1.4&0.108$\pm$0.009&7.76$\pm$1.62$^2$\\
R17e&765.061068&650.4&210&1.302&--1.4$\pm$0.7&15.4$\pm$2.5&0.051$\pm$0.006&6.77$\pm$1.40$^0$\\
R18e&767.997835&726.9&222&1.315&--0.7$\pm$0.5&7.9$\pm$1.9&0.068$\pm$0.012&4.70$\pm$1.57$^0$\\
\enddata
\tablecomments{See Table \ref{tab:HNClines} for column descriptions. Data in the first five columns are from the HITRAN database \citep{Gordon2017}. Superscripts in the final column refer to which velocity component $N_l$ will be totalled towards for the rotation diagrams in \S \ref{ssec:rot}: $^0$none, $^1$primary, $^2$secondary. Velocity for lines with more than one Gaussian is taken to be the that of the line with the largest $\tau_0$ and the other Gaussian is recorded without velocity and in italics.}
\vspace{-5mm}
\end{deluxetable*}

The LSR velocities of HNC, \hcniso, and the primary component of HCN fall in the range of $\sim -6$ to $-8$ \kms, making it highly likely that they trace the same component in IRc2. The secondary HCN velocity component is $\sim 1$ \kms. The line widths (\vfwhm $=2\sqrt{2\ln2}\sigma_v$) of HNC are systematically higher than the primary component of HCN by about 3 \kms.  Furthermore, \citet{Rangwala2018} observing the same HCN transitions as the present work, only resolve a single HCN velocity component with \vfwhm\  between 9 to 14 \kms, which is comparable to the HNC line widths in Tabe \ref{tab:HNClines}. Clearly, HNC may also have an unresolved, secondary velocity component.

Two HCN lines, R17e and R18e, are not counted towards any component, due to high noise and LSR velocities of $\sim-1$ \kms, which do not fit with either component. These lines were not resolved enough for a fit consistent with the other lines, but we include their derived parameters nonetheless in Table \ref{tab:HCNlines} for completeness as they are clearly detected in \citet{Rangwala2018}.

\subsection{Rotation Diagrams} \label{ssec:rot}

For molecular populations in local thermal equilibrium (LTE), the level populations follow a Boltzman distribution \citep{Goldsmith1999} given by:
\begin{equation}
    \ln \frac{N_l}{g_l}=\ln \frac{N}{Q_R(T_{ex})}-\frac{E_l}{k_b T_{ex}}
    \label{eqn:rot}
\end{equation}
where $N$ is the total column density, $Q_R$ is the rotational partition function, $E_l$ is the energy level of the lower state, and $T_{ex}$ is the excitation temperature. We calculated $Q_R$ for HCN and \hcniso\ for a given $T_{ex}$ with the HITRAN python interface HAPI \citep{Kochanov2016} and for HNC from levels published in ExoMol \citep{Harris2006,Barber2013}.

\begin{figure*}
\centering
\gridline{\fig{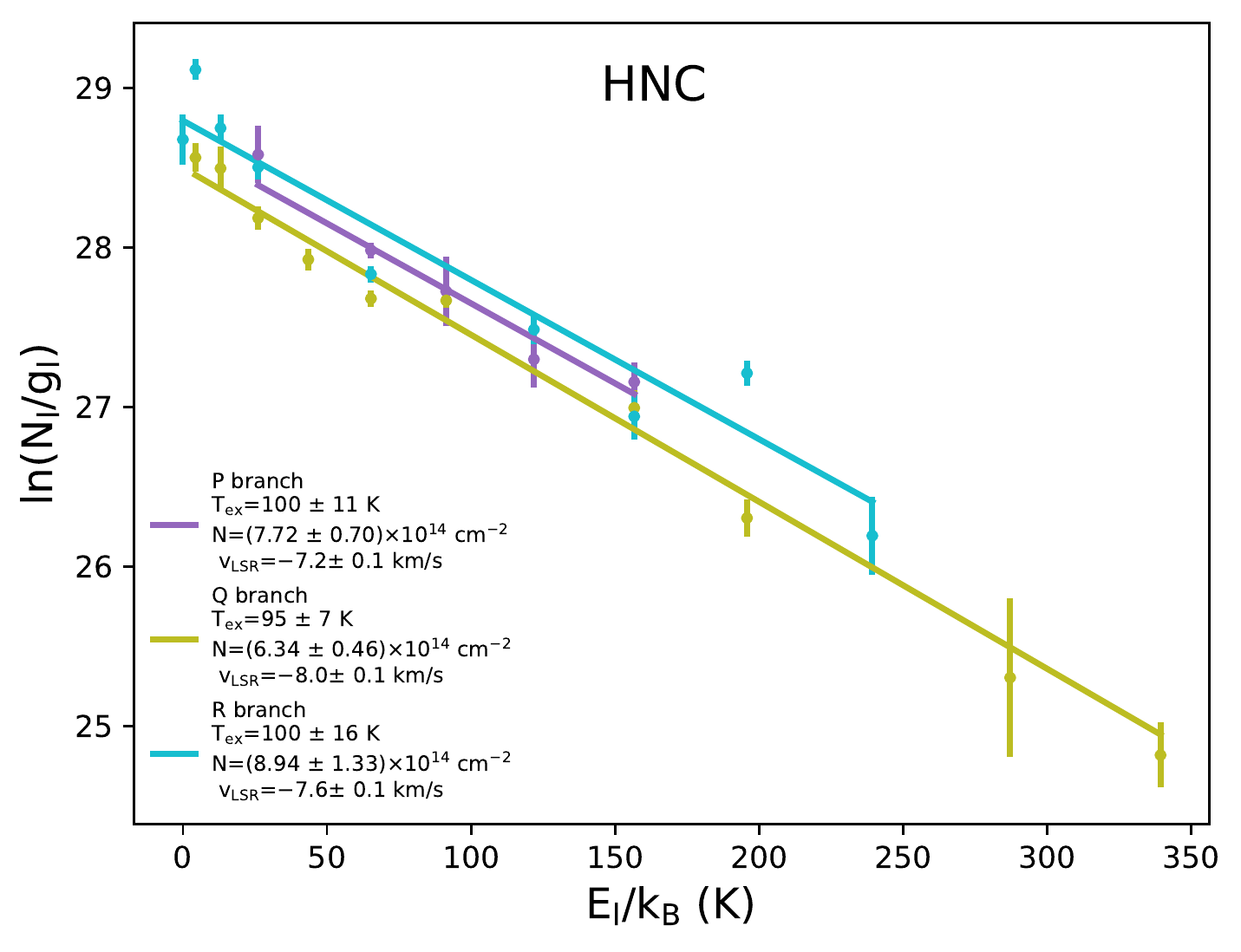}{0.49\textwidth}{}\hspace{-5mm}
          \fig{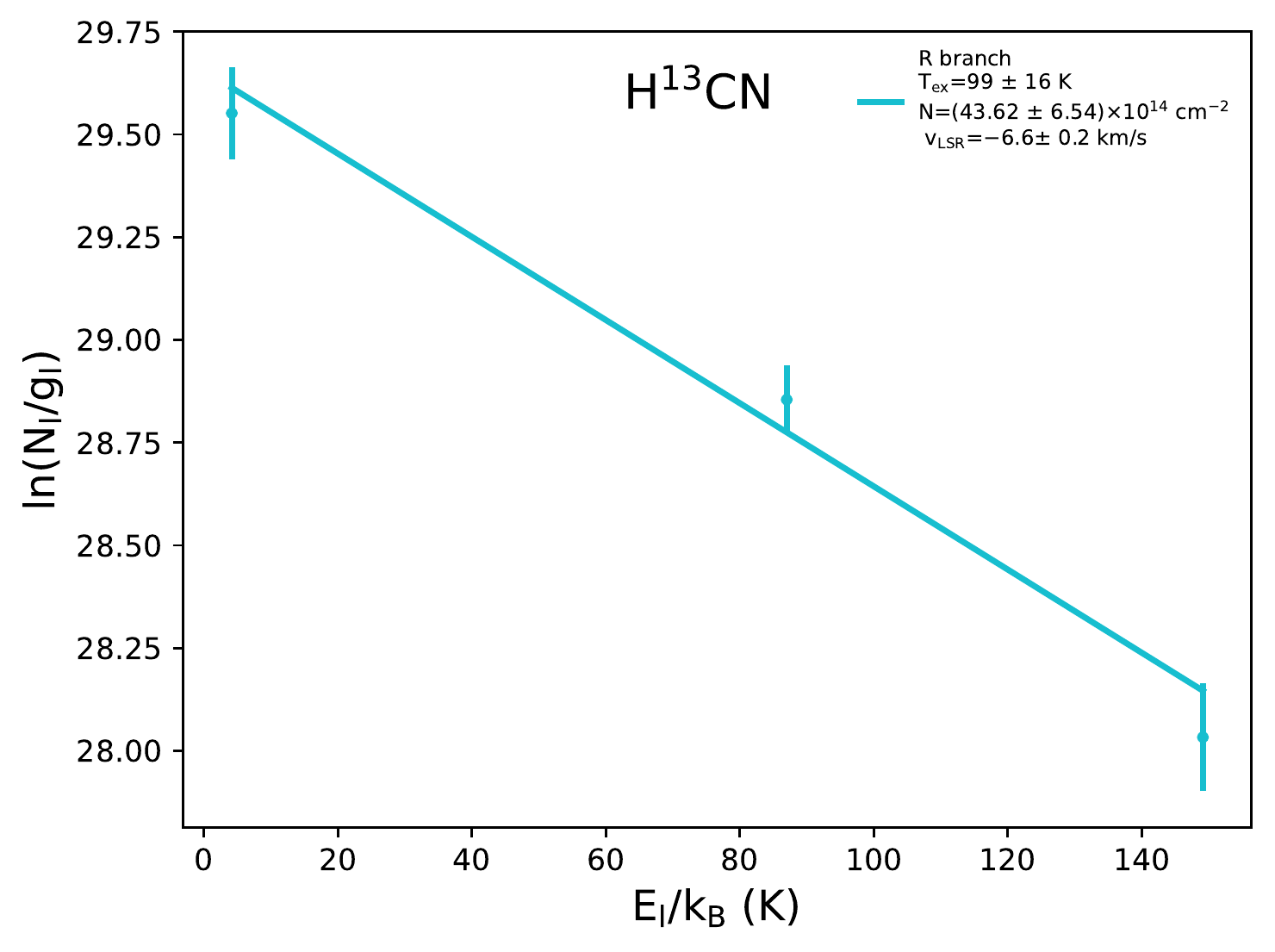}{0.49\textwidth}{}}
\vspace{-10mm}
\gridline{\fig{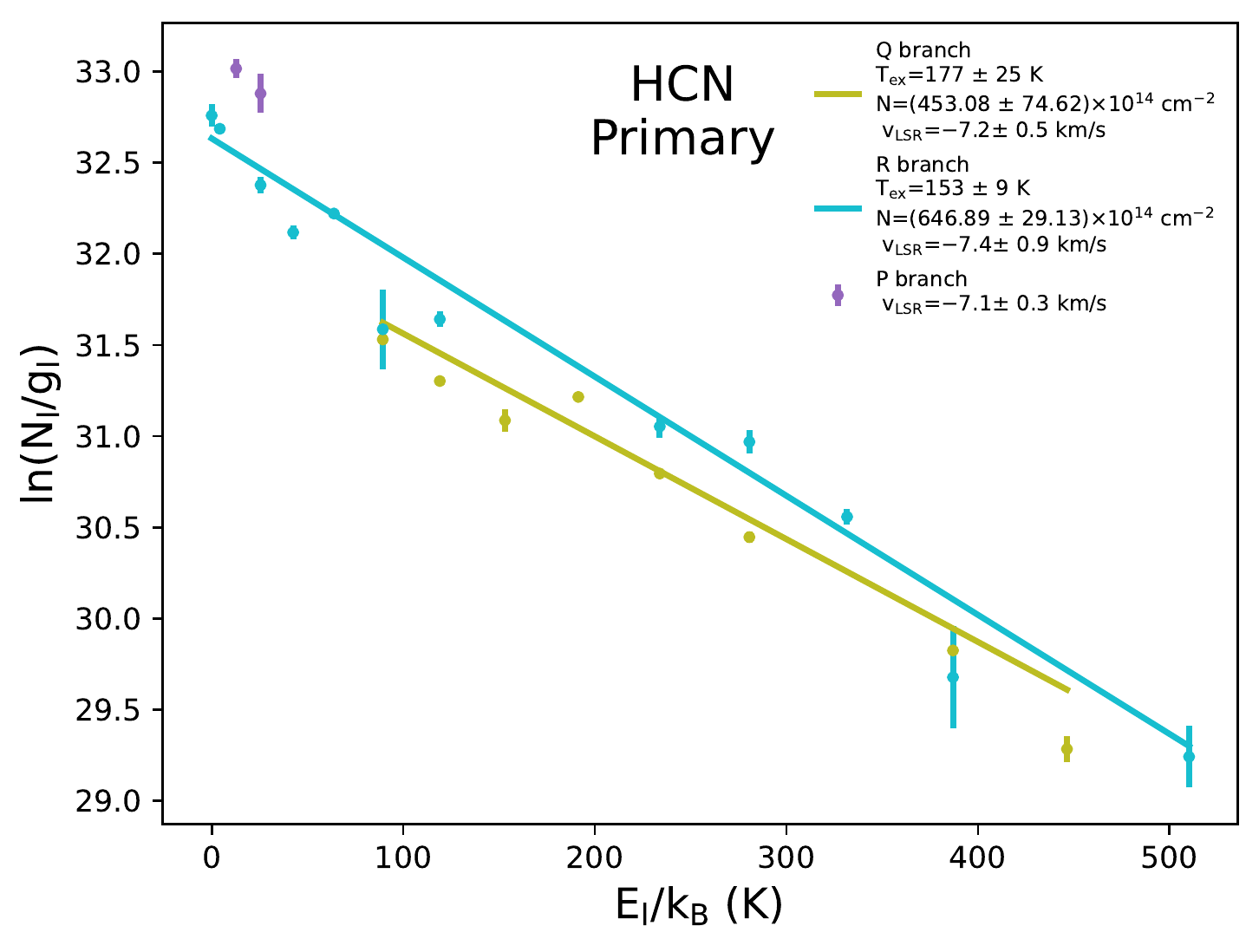}{0.49\textwidth}{}\hspace{-5mm}
          \fig{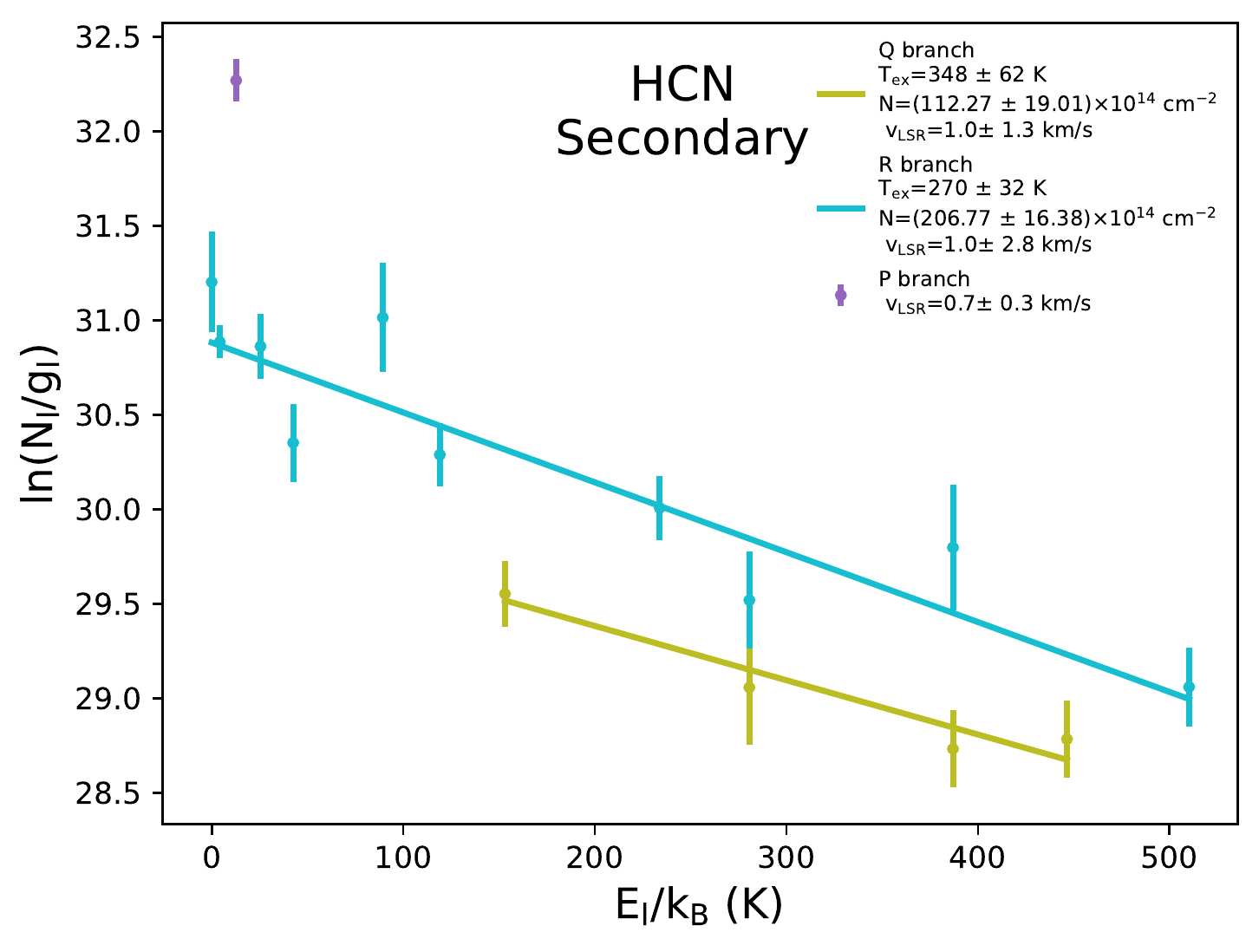}{0.49\textwidth}{}}
\vspace{-10mm}
\caption{Rotation diagrams for HNC (top left), \hcniso\ (top right), HCN primary velocity component (bottom left), and HCN secondary velocity component (bottom right). These follow Equation \ref{eqn:rot} where the excitation temperature and total column density of each species is extracted from the relation between the energy levels, column densities, and lower statistical weights of the transitions. The average LSR velocity for each branch is also given. \label{fig:rot}}
\end{figure*}

Figure \ref{fig:rot} gives the rotation diagrams for HNC, \hcniso, and the two velocity components of HCN. This shows the linear relationship between $\ln(N_l/g_l)$ and $E_l/k_b$, implying that the LTE approximation holds. By linearly fitting to Equation \ref{eqn:rot}, we obtain the values of $T_{ex}$ and $N$ for these species and components, summarized in Table \ref{tab:rot} along with their abundance relative to \htwo. 

Several observations give \nhtwo $\sim 10^{23}$ to $5\times10^{24}$ \csi\ towards IRc2 \citep[e.g.,][]{Evans1991,Schilke1992,Sutton1995,Persson2007,Tercero2010,Favre2011,Plume2012,Crockett2014,Feng2015} and we adopt the most recent calculation derived from ALMA dust continuum emission, \nhtwo$=(4.7\pm0.2)\times10^{24}$ \csi\ \citep{Peng2019}, as it has a comparable beam size to SOFIA/EXES. However, uncertainties stand with adopting this value because the \htwo\ distribution may differ from the that of the larger molecules HCN and HNC, and these observations may correspond to a different component of the ISM towards IRc2.  We do not fit the P branch of HCN, given only two points for the primary component and one for the secondary. Most of the P branch lines fell between observed settings.

\begin{deluxetable*}{ccccc}
\tablecaption{Overview of Species Properties\label{tab:rot}}
\tablehead{
\colhead{Branch} & \colhead{\vlsr} & \colhead{$T_{\mathrm{ex}}$} &
\colhead{$N$} & \colhead{$N/N_{\mathrm{H}_2}$} \\
\colhead{} & \colhead{(\kms)} & \colhead{(K)} &\colhead{$\times 10^{14}$ \csi}&\colhead{$\times 10^{-9}$}}
\startdata
&&HNC&&\\
\hline
P&--7.2$\pm$0.1&100$\pm$11&7.72$\pm$0.70&0.16$\pm$0.02\\
Q&--8.0$\pm$0.1&95$\pm$7&6.34$\pm$0.46&0.13$\pm$0.01\\
R&--7.6$\pm$0.1&100$\pm$16&8.94$\pm$1.33&0.19$\pm$0.03\\
\hline
&&HCN&&\\
\hline
P&--7.1$\pm$0.3&---&---&---\\
&0.7$\pm$0.3&---&---&---\\
Q&--7.2$\pm$0.5&177$\pm$25&453.08$\pm$74.62&9.64$\pm$1.64\\
&1.0$\pm$1.3&348$\pm$62&112.27$\pm$19.01&2.39$\pm$0.42\\
R&--7.4$\pm$0.9&153$\pm$9&646.89$\pm$29.13&13.76$\pm$0.85\\
&1.0$\pm$2.8&270$\pm$32&206.77$\pm$16.38&4.40$\pm$0.40\\
\hline
&&\hcniso&&\\
\hline
R&--6.6$\pm$0.2&99$\pm$16&43.62$\pm$6.54&0.93$\pm$0.14\\
\enddata
\tablecomments{For each species, branch, and velocity component, \vlsr\ is the average local standard rest of velocity, $T_{ex}$ is the excitation temperature, $N$ is the total column density, and \nhtwo\ is the column density of \htwo, adopted to be $4.7\pm0.2\times10^{24}$ \csi\ \citep{Peng2019}. The HCN P branch did not have enough points for a linear fit to find temperature and column density.}
\end{deluxetable*}

Both HNC and \hcniso\ trace the coldest gas at $\sim100$ K, while HNC has a lower column density at $7.7\times10^{14}$ \csi\ compared to $4.4\times10^{15}$ \csi\ for \hcniso. HCN traces the hottest, highest column density gas. Its primary velocity component (Q: 177 K; R: 153 K) is cooler than its secondary component (Q: 348 K; R: 270 K). However, the primary component has a higher column density (Q: $4.5\times10^{16}$ \csi; R: $6.5\times10^{16}$ \csi) compared to the secondary in both branches (Q: $1.1\times10^{16}$ \csi; R: $2.1\times10^{16}$ \csi).

\vspace{20mm}
\section{Discussion} \label{ssec:dis}

\subsection{Interpretation of Results} \label{ssec:interp}

The $-7$ \kms\ component of HNC, HCN, and \hcniso\ has also been observed in the MIR by \citet{Rangwala2018} and \citet{Lacy2005} for HCN and \acet, blueshifted with respect to the ambient cloud velocity of 9 \kms\ \citep{Zapata2012}. As discussed in \citet{Rangwala2018}, this velocity is similar to the North-East segment of the bipolar outflow originating from radio source I, which has been identified and mapped with SiO and \water\ masers \citep{Genzel1981,Wright1995,Plambeck2009}. Our $-7$ \kms\ lines may be associated with this outflow, being expelled material irradiated by IRc2, while the 1 \kms, hotter component is very likely closer to the hot core itself. Though we only resolved the 1 \kms\ component for HCN, the FWHMs of HNC in Table \ref{tab:HNClines} are systematically larger than the HCN lines in Table \ref{tab:HCNlines}. It may be that HNC also has a 1 \kms\ component that is unresolved by our observations. Our \hcniso\ lines are too weak for this analysis.

The HNC column densities in Table \ref{tab:rot} are consistently lower than HCN by about two orders of magnitude, evidence for a highly irradiated environment. Indeed, studies of planetary nebulae and the Helix Nebula find that HCN/HNC correlates with increasing UV radiation \citep{Bublitz2019,Bublitz2020}. HNC is less photostable than HCN \citep{Chenel2016,Aguado2017} and is destroyed in increasingly advanced and hotter stages of stellar evolution \citep{Jin2015}.  We see that towards Orion IRc2, HCN is more abundant and at hotter temperatures compared to HNC, in support of these findings.

A similar high resolution ($\sim$ 55,000 to 85,000) MIR spectral survey covered the hot cores AFGL 2591 and AFGL 2136 from 4 to 13 \micron\ with SOFIA/EXES, and the TEXES and the iSHELL instruments on the NASA Infrared Telescope Facility \citep{Barr2020}. In both hot cores, they identified transitions from the HCN  $\nu_2$ band, R branch that are also present in our Orion IRc2 survey. Their rotation diagrams for HCN yield temperatures $674.8\pm32.0$ K and $624.6\pm19$ K for AFGL 2591 and AFGL 2136 respectively. This is much hotter than the R branch temperatures for IRc2 that we find, $153\pm9$ K and $270\pm32$ K for each velocity component. IRc2 may be cooler than these two conventional hot cores due to its lack of an internal heat source. Similar to our conclusion, \citet{Barr2020} emphasize that the MIR probes hot gas at the centre of hot cores, compared to the sub-mm that probes the outer envelope.

\vspace{5mm}
\subsection{Comparison to Previous IRc2 Observations} \label{ssec:obs}

Table \ref{tab:comp} summarizes derived velocities, temperatures, and column densities for HNC, HCN, and \hcniso\ for observations towards Orion IRc2 from the literature and compares these to the present study. We take the average across all branches from Table \ref{tab:rot} for each component to display in Table \ref{tab:comp}. The most apparent difference is that the two spectral regimes reveal different components towards Orion IRc2. The sub-mm/mm observations detect these three species in emission with a positive \vlsr\  while the MIR observations detect them in absorption with a negative \vlsr.  Another difference is that all observations in the sub-mm/mm and MIR with \textit{ISO} \citep{Boonman2003} have a much larger beam size than EXES \citep{Rangwala2018} and TEXES \citep{Lacy2002,Lacy2005}. The smaller beam is important to observing the ISM component closest to the hot core. However, previous MIR observations do not resolve the secondary HCN velocity component present in our work \citep{Lacy2005,Rangwala2018}. Our MIR observations are able to probe closer to the host core itself, and find the hottest measured temperature for HCN towards Orion IRc2 to date.

Table \ref{tab:ratio} compares HCN/HNC and $^{12}$C/$^{13}$C ratios from this work to the literature. We use the average values over all branches of HCN, HNC, and \hcniso\ from Table \ref{tab:obs} for these ratios, taking only the  $-7$ \kms\ component of HCN, to find HCN/HNC=$72\pm7$ and $^{12}$C/$^{13}$C$=13\pm2$.

Our ratio HCN/HNC=$72\pm7$ falls within the the lower end of previous measurements that range from 75 to 300 \citep{Goldsmith1986,Schilke1992,Comito2005}. Such a range of values makes it difficult to draw direct a comparison between observations. Widely differing beam sizes may be responsible, where ours is the smallest at 3\farcs2, and others are: 11\arcsec\ \citep{Comito2005}, 13 to 26\arcsec\ \citep{Schilke1992}, and 42 to 60\arcsec\ \citep{Goldsmith1986}. We should also note that we measure a different velocity components than the sub-mm/mm studies, and that there are no previous studies in the MIR with which to compare our results directly. Our smaller beam size allows us to probe the hot core more closely.

The $^{12}$C/$^{13}$C$=13\pm2$ ratio towards IRc2 has been measured by more studies. Because we only saw three lines of \hcniso, the uncertainty in the ratio may be larger than quoted. With SOFIA/EXES MIR measurements of \acet, \citet{Rangwala2018} find $^{12}$C/$^{13}$C$=14\pm1$, similar to our value despite the difference in species to obtain it. Both these values are close to the \citet{Tercero2010} estimation for the hot core using H$_2$CS, $20\pm9$, but are much lower than other measurements that range from 53 to 82.6 \citep{Schilke1997,Favre2014,Feng2015}, using HCN, HCOOCH$_3$, and CH$_3$CN respectively. 

With Orion's distance from the Galactic Center and a linear relation derived from \citet{Milam2005}, \citet{Favre2014} calculated that the expected ratio is 50 to 90. This value is similar to measurements towards the neighboring compact ridge from 30 to 80 \citep{Blake1987,Persson2007,Gong2015}. One explanation for the lower ratio we and \citet{Rangwala2018} find towards the hot core is that previous studies  conducted in the sub-mm/mm and radio wavelengths observe different components than we do in the MIR, as illustrated in Table \ref{tab:comp}. As our lines are optically thin, then there must be either a physical or chemical explanation for this low ratio.

Our $^{12}$C/$^{13}$C is comparable to the ratio towards the Galactic Central hot core Sgr B2, $\sim 20$ \citep{Favre2014,Giesen2020}, which is expected for the region \citep{Milam2005,Yan2019}. Several measurements have found the ratio to be lower in star-forming regions than in the local ISM: 20-50 \citep{Daniel2013}, $\sim 30$ towards a protostellar binary \citep{Jorgensen2018}, 45 $\pm$ 3,\citep{Magalhaes2018}, and $\sim 16$ \citep{Bogelund2019} for various molecules. \citet{Colzi2020} suggest that an exchange reaction between $^{13}$C and C$_3$ can lead to a $^{13}$C enhancement in molecular clouds while they are still $<$30 K. This is cooler than our excitation temperatures for all species in Table \ref{tab:obs}. In this case, our low ratio may reflect conditions prior to protostellar warmup.

\begin{deluxetable*}{cccccccc}
\tablecaption{Comparison of HNC, HCN, and \hcniso\ observations towards Orion IRc2 between this and previously published works.\label{tab:comp}}
\tablehead{\colhead{Reference} & \colhead{Region}&\colhead{Beam Size}&\colhead{$\tau$}&\colhead{Type}& \colhead{\vlsr}&\colhead{$T$}&\colhead{$N$}\\\colhead{}&
\colhead{}&\colhead{}&\colhead{}&\colhead{}&\colhead{(\kms)}&\colhead{(K)}&\colhead{($\times 10^{14}$\csi)}}
\startdata
&&&HNC&&&\\
\hline
This work&MIR&5\farcs5--6\farcs9$\times$3\farcs2&thin&abs&$-7.6\pm$0.1&98$\pm$7&7.67$\pm$0.52\\
\citet{Persson2007}&sub-mm&2\farcm1&thin&emi&9&---&4.4\\
\citet{Comito2005}&sub-mm&11\arcsec&thin&emi&8&150\tablenotemark{a}&5\\
\hline
&&&HCN&&&\\
\hline
This work&MIR&1\farcs9--2\farcs1$\times$3\farcs2&thin&abs&$-7.3\pm$0.5&165$\pm$13&549.99$\pm$40.05\\
&&&&&1.0$\pm$1.5&309$\pm$35&159.52$\pm$12.55\\
\citet{Rangwala2018}&MIR&3\farcs2$\times$3\farcs2&thin&abs&$-5.2\pm2.8$&$140\pm10$\tablenotemark{b}&$840\pm60$\\
\citet{Comito2005}&sub-mm&11\arcsec&thin&emi&5&150\tablenotemark{a}&700\\
\citet{Lacy2005}&MIR&1\farcs5$\times$8\arcsec\tablenotemark{c}&---\tablenotemark{c}&abs&$-10\tablenotemark{d}$&150\tablenotemark{d}&---\\
\citet{Boonman2003}&MIR&14\arcsec$\times$27\arcsec&thick&abs&---&275\tablenotemark{e}&450\\
\citet{Schilke2001}&sub-mm&12\arcsec&thin&emi&---&100\tablenotemark{f}&260\\
\citet{Stutzki1988}&sub-mm&32\arcsec&thick&emi&6&---&---\\
\citet{Blake1987}&mm&30\arcsec&thick&emi&5.8&---&250\\
\hline
&&&\hcniso&&&\\
\hline
This work&MIR&2\farcs1$\times$3\farcs2&thin&abs&$-6.6\pm$0.2&97$\pm$13&42.08$\pm$5.67\\
\citet{Schilke1992}&mm&26\arcsec&thin&emi&9.2\tablenotemark{g}&70\tablenotemark{g}&0.57\\     
\enddata
\tablecomments{Left to right, columns refer to observation reference, spectral region, beam size, whether the observation is optically thin or thick, whether the line type is emission (emi) or absorption (abs), LSR velocity, temperature, and column density. Temperatures are calculated by a variety of methods, as detained in table footnotes. The HNC and HCN values in this work are the averages across all branches for each component from Table \ref{tab:rot}. For reference, the ambient cloud velocity is 9 \kms\ \citep{Zapata2012}. \tablenotetext{a}{rotation temperature derived from DCN}\tablenotetext{b}{excitation temperature, R branch only} \tablenotetext{c}{beam size from \citet{Lacy2002}, opacity unknown}\tablenotetext{d}{$v_{\mathrm{LSR}}$ and temperature approximation from HCN and \acet\ lines}\tablenotetext{e}{modelled excitation temperature}\tablenotetext{f}{rotation temperature derived from H$^{15}$CN} \tablenotetext{g}{$v_{\mathrm{LSR}}$ and kinetic temperature estimated for entire hot core}}
\end{deluxetable*}

\begin{deluxetable}{ccc}
\tablecaption{Comparison of observed ratios towards Orion IRc2\label{tab:ratio}}
\tablehead{\colhead{Reference}&Spectral Region&\colhead{Ratio}}
\startdata
&HCN/HNC&\\
\hline
This work&MIR&$72\pm7$\\
\citet{Comito2005}&sub-mm&140\\
\citet{Schilke1992}&mm&$80\pm5$\\
\citet{Goldsmith1986}&mm&$\sim$200--300\\
\hline
&$^{12}$C/$^{13}$C&\\
\hline
This work&MIR&$13\pm2$\tablenotemark{a}\\
\citet{Rangwala2018}&MIR&$14\pm1$\tablenotemark{b}\\
\citet{Feng2015}&sub-mm&$79.6\pm3.0$\tablenotemark{c}\\
\citet{Favre2014}&mm&$\geq$53\tablenotemark{d}\\
\citet{Tercero2010}&mm&$20\pm9$\tablenotemark{e}\\
\citet{Schilke1997}&radio&60\tablenotemark{a}\\
\enddata
\tablecomments{Species used to obtain $^{12}$C/$^{13}$C ratio: \tablenotetext{a}{HCN}\tablenotetext{b}{\acet}\tablenotetext{c}{CH$_3$CN}\tablenotetext{d}{HCOOCH$_3$}\tablenotetext{e}{H$_2$CS}}
\end{deluxetable}

\vspace{6mm}
\subsection{Comparison to a Hot Core Model} \label{ssec:model}

\begin{figure}
\centering
   \includegraphics[width=0.44\textwidth]{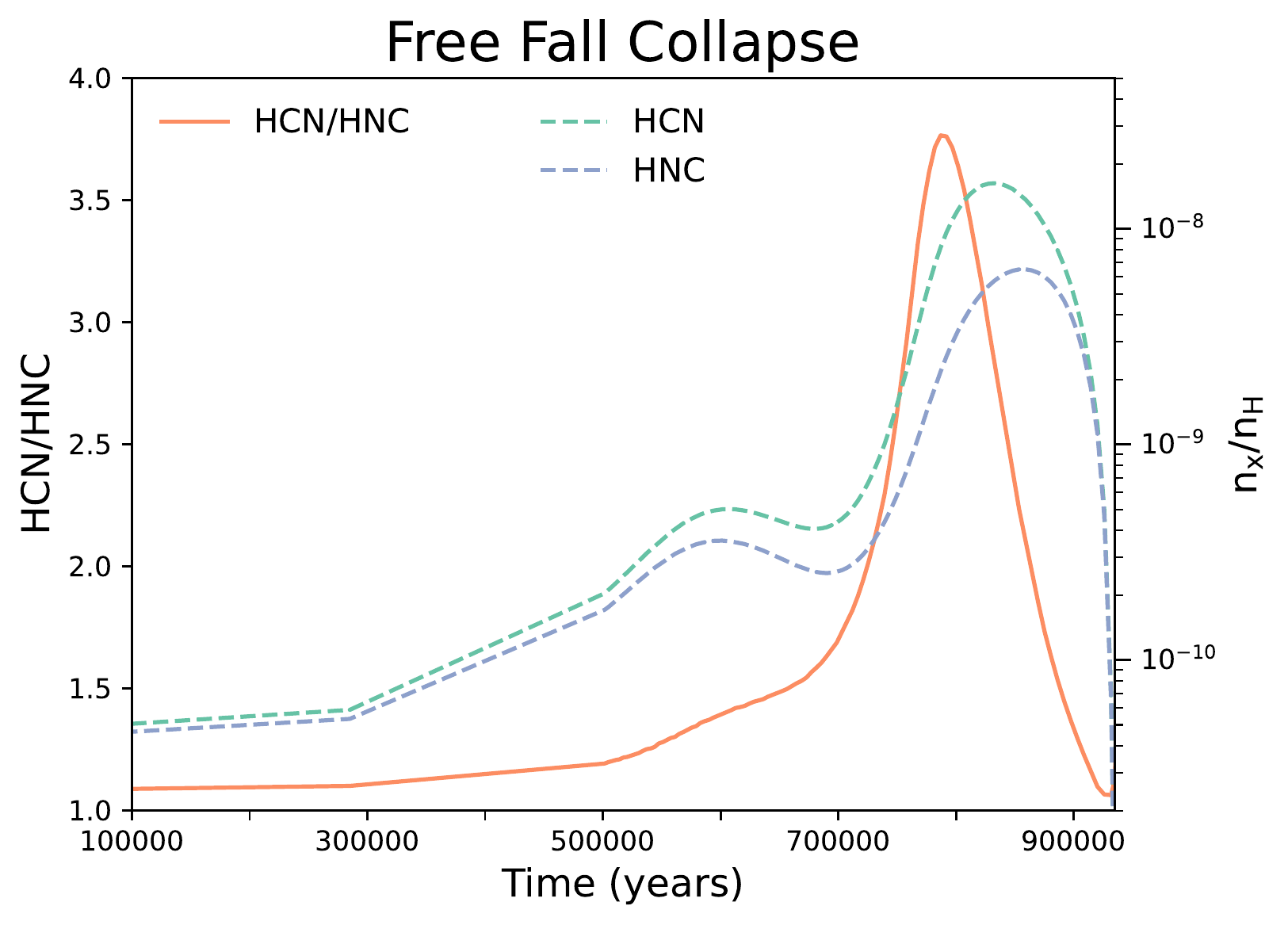}
\caption{Time variation of HCN/HNC (orange solid line, left axis) and HCN and HNC abundances (green and blue dotted lines, right axis) during the free fall collapse phase of the model.}
\label{fig:freefall}
\end{figure}

\begin{figure*}
\centering
   \includegraphics[width=0.92\textwidth]{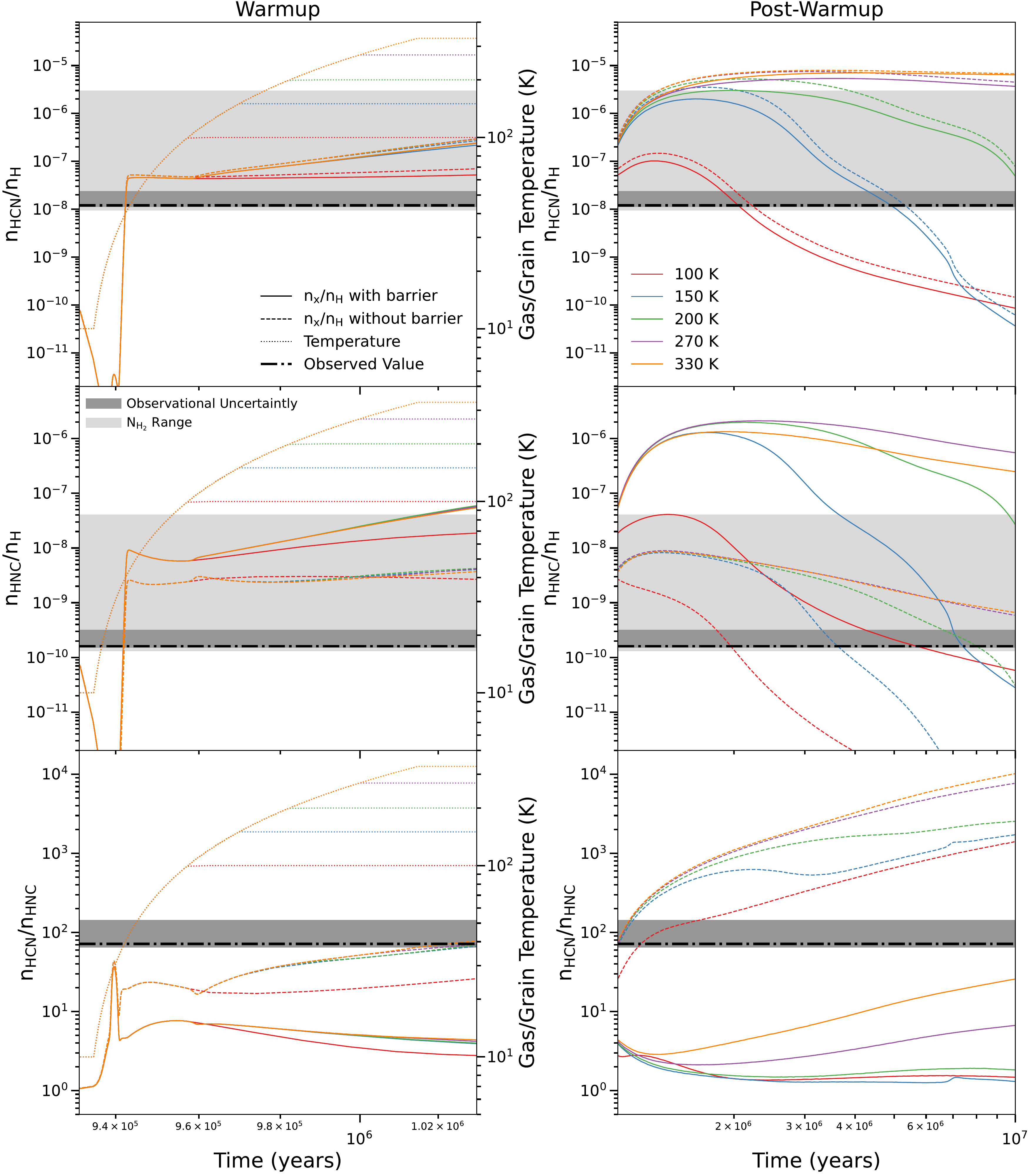}
\caption{Time variation of HCN (top tow), HNC (middle row), and HCN/HNC (bottom row) during the warmup (left column) and post-warmup phase (right column) on the left y-axes. Solid lines and dashed lines correspond to HNC + H $\rightarrow$ HCN + H  with and without a 2000 K barrier respectively. Time variation of gas/grain temperature is shown by dotted lines corresponding to the right y-axis during the warmup phase. Colour for all three line types corresponds to the final temperature reached during the warmup phase. The horizontal dash-dotted lines represent the observed values for $N_{\mathrm{HCN}}/N_{\mathrm{H}_2}$, $N_{\mathrm{HNC}}/N_{\mathrm{H}_2}$, and $N_{\mathrm{HCN}}/N_{\mathrm{HNC}}$ derived from Table \ref{tab:rot} by taking the average across all branches for the $-7$ \kms\ component and using our adopted \nhtwo$=(4.7\pm0.2)\times10^{24}$ \csi\ \citep{Peng2019}. The dark grey box corresponds to the error. The light grey box reflects the extrema in the measured N$_{\mathrm{H}_2}$ considering the lowest value 2$\times10^{22}$\csi\ from \citet{Persson2007} and the highest value of 5.4$\times10^{24}$\csi\ from \citet{Favre2011}. Legends apply to all the panels.}
\label{fig:warmup}
\end{figure*}

To compare our observations to modelling, we utilized the gas-grain chemical network as described in \citet{Acharyya2018} for the model in Figures \ref{fig:freefall} and \ref{fig:warmup}. Both the gas-phase and grain-surface chemistry are treated via a rate equation approach. The major features of the model are as follows:
\begin{itemize}
\item Model runs have two physical evolutionary phases as prescribed by \citet{Brown1988}. In the first phase, free fall collapse (Figure \ref{fig:freefall}), the cloud undergoes isothermal collapse at 10 K, from a density of 3,000 to 10$^7$ \cci\ in just under 10$^6$ years. During this period the visual extinction grows from 1.64 to 432 mag. In the second phase, warmup (Figure \ref{fig:warmup}, left column), the collapse stops and the temperature increases linearly at the rate of 1 K/250 years, which is a representative heating rate for  high-mass star-formation. We ran warmup models for five different final temperatures: 100, 150, 200, 270 and 300 K. Once the hot core reaches its final temperature, its chemical evolution continues through the third phase, post-warmup, until 10$^7$ years elapse (Figure \ref{fig:warmup}, right column).
\item \citet{Graninger2014} found that the reaction HNC + H $\rightarrow$ HCN + H regulates the HCN/HNC ratio under all conditions. However, the activation barrier of this reaction is highly uncertain, and can vary between 0 and 2,000 K. Therefore, we study the cases with and without the 2,000 K activation barrier.
\item Classical dust grains are 0.1 \micron\ in size with a surface site density $n_{\rm s}$ = 1.5 $\times$ 10$^{15}$ \csi, leading to about 10$^6$ adsorption binding sites per grain.
\item We use standard low-metal elemental abundances, initially in the form of gaseous atoms with the exception of \htwo. Elements with ionization potentials below 13.6 eV take the form of singly charged positive ions (C$^+$, Fe$^+$, Na$^+$, Mg$^+$, S$^+$, Si$^+$, and Cl$^+$).
\item We use a sticking coefficient of 1, standard cosmic-ray ionisation rate of 10$^{-17}$ s$^{-1}$, and a diffusion-to-binding energy ratio of 0.5.
\end{itemize}

Figure \ref{fig:freefall} shows the time variation of HCN and HNC abundances 
during the freefall collapse phase during which the temperature is constant at 10 K. The peak abundances  (relative to total atmoic hydrogen) for HCN and HNC are 1.6 $\times$ 10$^{-8}$ and 6.5 $\times$ 10$^{-8}$ respectively. Towards the end of the first phase, HCN and HNC freeze out on the dust grains. The peak HCN/HNC ratio is $\sim$ 3.8. 

Figure \ref{fig:warmup} shows the time variation of HCN and HNC for ten different models during the warmup and post-warmup phases, with five different maximum temperatures of 100, 150, 200, 270 and 330 K, with or without the activation barrier. The left column of Figure \ref{fig:warmup} shows this chemical evolution during the warmup phase and the right column shows this in the post-warmup phase. Usually, the warmup phase lasts only up to $\sim$ 10$^4$ years. However, we have plotted up to 10$^7$ years to show the effect of a long lifetime. 

It is clear that for HCN the abundances for all models are very similar until the warmup phase, with the exception of the 100 K model. However in the post-warmup phase, models deviate slightly depending on activation barrier, and varying temperature causes more deviation. For HNC, the deviation with and without the activation barrier begins in the warmup phase by a factor of 20, and all ten models deviate the most during the post-warmup phase. Overall, the differences between models during the warmup phase are small, and they vary more during the post-warmup phase. 

During post-warmup, the abundances for HCN in the 100 and 150 K models decreases by several orders of magnitude; whereas for the 200 K model, the reduction is about two orders of magnitude; and for the 270 and 330 K models, the change is minor. For HNC, a large reduction is present for all the models with no activation barrier. With a barrier, only the 100 and 150 K models show a large reduction.

Figure \ref{fig:warmup} also includes lines for $N_{\mathrm{HCN}}/N_{\mathrm{H}_2}$, $N_{\mathrm{HNC}}/N_{\mathrm{H}_2}$, and $N_{\mathrm{HCN}}/N_{\mathrm{HNC}}$ as measured from Table \ref{tab:rot} taking the average across all branches for the $-7$ \kms\ velocity component. We do not display the 1 \kms\ component of HCN given that there is no similar component in HNC. Our observed values for $N_{\mathrm{x}}/N_{\mathrm{H}_2}$ are comparable to the simulation's $n_{\mathrm{x}}/n_{\mathrm{H}}$ because in this situation we can assume that all hydrogen is molecular. We take \nhtwo$=(4.7\pm0.2)\times10^{24}$ \csi\ \citep{Peng2019} as our default value from the literature, given the similarity in the ALMA beam size to EXES. However, as discussed in \S \ref{ssec:rot}, the observed \nhtwo\ towards IRc2 varies greatly. We also take the extreme values from 2$\times10^{22}$ \csi\ \citep{Persson2007} to 5.4$\times10^{24}$ \csi\ \citep{Favre2011}. This shows how much the uncertainty in our HCN and HNC column densities is dwarfed by the range of values that \nhtwo\ takes in the literature. Because of this, $N_{\mathrm{HCN}}/N_{\mathrm{HNC}}$ offers a much more robust comparison to the model.
 
Indeed, our HCN and HNC abundances fall short of all ten models employed during the warmup phase. They coincide slightly for the 100 K no-barrier model at $2\times10^6$ years, as well as the 150 K barrier model at about $7\times10^6$ years. If we take into consideration, however, that our abundances are dependent on the widely varying \nhtwo, then it is impossible to match our data to any one model as they all fall within the uncertainties. We prefer instead to make our prediction based on HCN/HNC, which does not require the molecular hydrogen column density. In this situation, our observations match the 150 to 330 K no-barrier models at about $10^6$ years. This is similar to the hot core age \citet{Rangwala2018} derived by matching their observed \acet\ and HCN abundances to hot core models.

The OMC-1 region was the site of a powerful explosion about 500 years ago \citep{Gomez2008}. The $10^{6}$ years required to produce HCN/HNC=$72\pm7$ is much longer than the 500 year-old explosive event in the Orion Molecular Cloud. Given the lack of evidence for an embedded protostar that internally heats IRc2, there are two explanations for this mismatch in timescales. The first is that IRc2 may have been irradiated prior to the explosive event by proximity to radio source I and was later expelled. As such, it would have resembled a traditional, internally heated hot core prior to explosion. 

The second possible explanation is that our modelling does not take shock physics into account, and the inclusion of such may alter the timescales needed to reach our HCN/HNC ratio and our predicted age. Recent observations \citep{Wright2017} suggest the region is indeed shocked. The competing role of C-type shocks in the Galactic Central hot core Sgr B2(N) has just been studied by \citet{Zhang2020}.  In addition, \citet{Willis2020} report new hot core models of Sgr B2(N) that use a single phase in which density and temperature are varied simultaneously in a study of isocyanides and cyanides, including HNC and HCN.

\section{Conclusions} \label{sec:conc}
From SOFIA/EXES we present high spectral resolution, MIR absorption spectra of HCN, HNC, and \hcniso\ towards the hot core Orion IRc2. This is the first MIR observations of HNC and \hcniso\  in the ISM. Almost continuous coverage of the P, Q, and R branch transitions of HNC, and the Q and R branch transitions of HCN yield detailed rotation diagrams that produce the species' excitation temperatures and column densities. Only three \hcniso\ R branch transitions are strong enough to include in a rotation diagram. All three species have a LSR velocity $-7$ \kms\ with average temperatures and column densities for HNC of 98 K and $7.7\times10^{14}$ \csi; \hcniso\ 97 K and $4.2\times10^{15}$ \csi; and HCN 165 K and $5.5\times10^{16}$  \csi. For the first time, we observe a second MIR velocity component of HCN measured at $1$ \kms, 309 K, and $1.6\times10^{16}$ \csi.

Our $-7$ \kms\ absorption lines belong to the same component measured in the MIR towards Orion IRc2 by \citet{Rangwala2018} and \citet{Lacy2005}, which probes different material than the emission lines observed in the sub-mm/mm \citep{Persson2007,Comito2005,Schilke2001,Schilke1992,Stutzki1988,Blake1987}. It is clear that the MIR is key to studying the ISM closest to the hot core compared to longer wavelength observations in sub-mm/mm. This $-7$ \kms\ component is similar in velocity to a bipolar outflow originating from nearby radio source I, which has no infrared component. Our HCN, HNC, and  \hcniso\ are likely associated with this outflow. 

The HCN in the 1 \kms\ component is much hotter and because of this, represents material closer to the hot core itself. EXES's small beam size compared to previous MIR and most grund-based sub-mm/mm instruments allows us to isolate the material at the hot core's centre. While HNC only has one clear velocity component, its line widths are consistently wider than HCN, hinting that it may have two components, albeit unresolved. We would expect HCN/HNC to be even higher in this hotter component. The \hcniso\ lines do not have a high enough signal-to-noise to comment on the possibly of a second component. 

We find HCN/HNC $=72\pm7$ in range of sub-mm/mm observations towards Orion IRc2 \citep{Schilke1992,Goldsmith1986}. This supports findings that this ratio is enhanced in hot, irradiated environments. Our isotopic ratio $^{12}$C/$^{13}$C$=13\pm2$ is lower than expected for Orion IRc2's galactocentric distance, though similar to measurements by \citet{Rangwala2018} in the MIR and \citet{Tercero2010} in the mm.

In order to determine the age of the hot core, we run the chemical network of \citet{Acharyya2018} for a hot core model with three phases: free fall collapse, warmup, and post-warmup. Comparison with HCN and HNC abundances alone are highly uncertain given the range in observed \nhtwo\ towards IRc2. Instead, we use our observed HCN/HNC ratio to find that the hot core is $\sim10^{6}$ years old. This supports similar results in \citet{Rangwala2018}.

This work  demonstrates the importance of the MIR in accessing transitions of HCN, HNC, and \hcniso\ that  originate in hotter material closer to the hot core compared to commonly used rotational transitions in the sub-mm/mm. The spectra presented here are a part of a larger survey with SOFIA/EXES towards Orion IRc2. Future papers detailing more molecular species are forthcoming. These studies will be critical in informing the observations from the James Webb Space Telescope (JWST), which while more sensitive, lacks the resolving power of SOFIA/EXES. With SOFIA/EXES we can unambiguously identify the signals from the strongest molecular species and construct a reference database to inform the search for weaker molecular species in JWST spectra. 

\acknowledgments
This work made use of the following Python packages: \texttt{Astropy} \citep{Price-Whelan2018,Robitaille2013}, \texttt{Matplotlib} \citep{Hunter2007}, \texttt{Numpy} \citep{VanDerWalt2011}, \texttt{Scipy} \citep{Virtanen2020}, and HAPI \citep{Kochanov2016}. We would also like to thank James De Buizer for providing FORCAST data, and Kyle Kaplan for the useful conversation. SN gratefully acknowledges funding from the
BAERI cooperative agreement 80NSSC17M0014. TJL, and XH gratefully acknowledge financial support from the NASA 16-PDART16\_2-0 080, 17-APRA17-0051, and 18-APRA18-0013 grants. XH acknowledges the NASA/SETI Co-operative Agreements NNX15AF45A, NNX17AL03G and 80NSSC19M0121. MND is supported by the Swiss National Science Foundation (SNSF) Ambizione grant 180079, the Center for Space and Habitability (CSH) Fellowship, and the IAU Gruber Foundation Fellowship. RCF acknowledges funding from NASA grant NNX17AH15G.

\bibliography{mendeley,slnrefs}{}

\begin{thebibliography}{}
\expandafter\ifx\csname natexlab\endcsname\relax\def\natexlab#1{#1}\fi
\providecommand{\url}[1]{\href{#1}{#1}}
\providecommand{\dodoi}[1]{doi:~\href{http://doi.org/#1}{\nolinkurl{#1}}}
\providecommand{\doeprint}[1]{\href{http://ascl.net/#1}{\nolinkurl{http://ascl.net/#1}}}
\providecommand{\doarXiv}[1]{\href{https://arxiv.org/abs/#1}{\nolinkurl{https://arxiv.org/abs/#1}}}

\bibitem[{Aalto {et~al.}(2012)Aalto, Garcia-Burillo, Muller, Winters, {Van Der
  Werf}, Henkel, Costagliola, \& Neri}]{Aalto2012}
Aalto, S., Garcia-Burillo, S., Muller, S., {et~al.} 2012, A{\&}A, 537, 1,
  \dodoi{10.1051/0004-6361/201117919}

\bibitem[{Acharyya \& Herbst(2018)}]{Acharyya2018}
Acharyya, K., \& Herbst, E. 2018, ApJ, 859, 51,
  \dodoi{10.3847/1538-4357/aabaf2}

\bibitem[{Aguado {et~al.}(2017)Aguado, Roncero, Zanchet, Ag{\'{u}}ndez, \&
  Cernicharo}]{Aguado2017}
Aguado, A., Roncero, O., Zanchet, A., Ag{\'{u}}ndez, M., \& Cernicharo, J.
  2017, ApJ, 838, 33, \dodoi{10.3847/1538-4357/aa63ee}

\bibitem[{Ag{\'{u}}ndez {et~al.}(2014)Ag{\'{u}}ndez, Biver, Santos-Sanz,
  Bockel{\'{e}}e-Morvan, \& Moreno}]{Agundez2014}
Ag{\'{u}}ndez, M., Biver, N., Santos-Sanz, P., Bockel{\'{e}}e-Morvan, D., \&
  Moreno, R. 2014, A{\&}A, 564, L2, \dodoi{10.1051/0004-6361/201423639}

\bibitem[{An {et~al.}(2009)An, Ram{\'{i}}rez, Sellgren, Arendt, Boogert,
  Schultheis, Stolovy, Cotera, Robitaille, \& Smith}]{An2009}
An, D., Ram{\'{i}}rez, S.~V., Sellgren, K., {et~al.} 2009, ApJ, 702, 128,
  \dodoi{10.1088/0004-637X/702/2/L128}

\bibitem[{An {et~al.}(2011)An, Ram{\'{i}}rez, Sellgren, Arendt, Boogert,
  Robitaille, Schultheis, Cotera, Smith, \& Stolovy}]{An2011}
---. 2011, ApJ, 736, 133, \dodoi{10.1088/0004-637X/736/2/133}

\bibitem[{Bally {et~al.}(2017)Bally, Ginsburg, Arce, Eisner, Youngblood,
  Zapata, \& Zinnecker}]{Bally2017}
Bally, J., Ginsburg, A., Arce, H., {et~al.} 2017, ApJ, 837, 60,
  \dodoi{10.3847/1538-4357/aa5c8b}

\bibitem[{Barber {et~al.}(2013)Barber, Strange, Hill, Polyansky, Mellau,
  Yurchenko, \& Tennyson}]{Barber2013}
Barber, R.~J., Strange, J.~K., Hill, C., {et~al.} 2013, Monthly Notices of the
  Royal Astronomical Society, 437, 1828, \dodoi{10.1093/mnras/stt2011}

\bibitem[{Barr {et~al.}(2018)Barr, Boogert, Dewitt, Montiel, Richter, Indriolo,
  Neufeld, Pendleton, Chiar, Dungee, \& Tielens}]{Barr2018}
Barr, A.~G., Boogert, A., Dewitt, C.~N., {et~al.} 2018, ApJ: Letters, 868, L2,
  \dodoi{10.3847/2041-8213/aaeb23}

\bibitem[{Barr {et~al.}(2020)Barr, Boogert, DeWitt, Montiel, Richter, Lacy,
  Neufeld, Indriolo, Pendleton, Chiar, \& Tielens}]{Barr2020}
Barr, A.~G., Boogert, A., DeWitt, C.~N., {et~al.} 2020, arXiv e-prints.
\newblock \doarXiv{2007.11266}

\bibitem[{Belloche {et~al.}(2013)Belloche, M{\"{u}}ller, Menten, Schilke, \&
  Comito}]{Belloche2013}
Belloche, A., M{\"{u}}ller, H.~S., Menten, K.~M., Schilke, P., \& Comito, C.
  2013, A{\&}A, 559, \dodoi{10.1051/0004-6361/201321096}

\bibitem[{Benson \& Myers(1989)}]{Benson1989}
Benson, P.~J., \& Myers, P.~C. 1989, ApJS, 71, 89

\bibitem[{Bisschop {et~al.}(2007)Bisschop, J{\o}rgensen, van Dishoeck, \& {De
  Wachter}}]{Bisschop2007}
Bisschop, S.~E., J{\o}rgensen, J.~K., van Dishoeck, E.~F., \& {De Wachter},
  E.~B. 2007, A{\&}A, 465, 913, \dodoi{10.1051/0004-6361:20065963}

\bibitem[{Blake {et~al.}(1996)Blake, Mundy, Carlstrom, Padin, Scott, Scoville,
  \& Woody}]{Blake1996}
Blake, G.~A., Mundy, L.~G., Carlstrom, J.~E., {et~al.} 1996, ApJ: Letters, 472,
  L49

\bibitem[{Blake {et~al.}(1987)Blake, Sutton, Masson, \& Phillips}]{Blake1987}
Blake, G.~A., Sutton, E.~C., Masson, C.~R., \& Phillips, T.~G. 1987, ApJ, 315,
  612

\bibitem[{B{\o}gelund {et~al.}(2019)B{\o}gelund, Barr, Taquet, Ligterink,
  Persson, Hogerheijde, \& van Dishoeck}]{Bogelund2019}
B{\o}gelund, E.~G., Barr, A.~G., Taquet, V., {et~al.} 2019, A{\&}A, 628, A2,
  \dodoi{10.1051/0004-6361/201834527}

\bibitem[{Boonman {et~al.}(2003)Boonman, van Dishoeck, Lahuis, Doty, Wright, \&
  Rosenthal}]{Boonman2003}
Boonman, A.~M., van Dishoeck, E.~F., Lahuis, F., {et~al.} 2003, A{\&}A, 399,
  1047, \dodoi{10.1051/0004-6361:20021799}

\bibitem[{Boonman {et~al.}(2001)Boonman, Stark, van~der Tak, van Dishoeck,
  van~der Wal, Sch{\"{a}}fer, de~Lange, \& Laauwen}]{Boonman2001}
Boonman, A. M.~S., Stark, R., van~der Tak, F. F.~S., {et~al.} 2001, ApJ, 553,
  L63, \dodoi{10.1086/320493}

\bibitem[{Bottinelli {et~al.}(2004)Bottinelli, Ceccarelli, Lefloch, Williams,
  Castets, Caux, Cazaux, Maret, Parise, \& Tielens}]{Bottinelli2004}
Bottinelli, S., Ceccarelli, C., Lefloch, B., {et~al.} 2004, ApJ, 615, 354,
  \dodoi{10.1086/423952}

\bibitem[{Bowman {et~al.}(1993)Bowman, Gazdy, Bentley, Lee, \&
  Dateo}]{Bowman1993}
Bowman, J.~M., Gazdy, B., Bentley, J.~A., Lee, T.~J., \& Dateo, C.~E. 1993, The
  Journal of Chemical Physics, 99, 308, \dodoi{10.1063/1.465809}

\bibitem[{Brown {et~al.}(1988)Brown, Charnley, \& Millar}]{Brown1988}
Brown, P.~D., Charnley, S.~B., \& Millar, T.~J. 1988, MNRAS, 231, 409

\bibitem[{Bublitz {et~al.}(2020)Bublitz, Kastner, Hily-Blant, Forveille,
  Santander-Garc{\'{i}}a, Bujarrabal, Alcolea, \& Montez}]{Bublitz2020}
Bublitz, J., Kastner, J., Hily-Blant, P., {et~al.} 2020, Galaxies, 8, 32,
  \dodoi{10.3390/galaxies8020032}

\bibitem[{Bublitz {et~al.}(2019)Bublitz, Kastner, Santander-Garc{\'{i}}a,
  Bujarrabal, Alcolea, \& Montez}]{Bublitz2019}
Bublitz, J., Kastner, J.~H., Santander-Garc{\'{i}}a, M., {et~al.} 2019, A{\&}A,
  625, 1, \dodoi{10.1051/0004-6361/201834408}

\bibitem[{Bujarrabal {et~al.}(1994)Bujarrabal, Fuente, \&
  Omont}]{Bujarrabal1994}
Bujarrabal, V., Fuente, A., \& Omont, A. 1994, A{\&}A, 285, 247

\bibitem[{Cernicharo {et~al.}(2013)Cernicharo, Daniel, Castro-Carrizo, Agundez,
  Marcelino, Joblin, Goicoechea, \& Gu{\'{e}}lin}]{Cernicharo2013}
Cernicharo, J., Daniel, F., Castro-Carrizo, A., {et~al.} 2013, ApJ: Letters,
  778, L25, \dodoi{10.1088/2041-8205/778/2/L25}

\bibitem[{Cernicharo {et~al.}(1996)Cernicharo, Barlow, Gonzalez-Alfonso, Cox,
  Clegg, Nguyen-Q-Rieu, Omont, Guelin, Liu, Sylvester, Lim, Griffin, Swinyard,
  Unger, Ade, Baluteau, Caux, Cohen, Emery, Fischer, I., Glencross, Greenhouse,
  Gry, Joubert, Lorenzetti, Nisini, Orfei, Pequignot, Saraceno, Serra, Skinner,
  Smith, Towlson, Walker, Armand, Burgdorf, Ewart, di~Giorgio, Molinari, Price,
  Sidher, Texier, \& Trams}]{Cernicharo1996}
Cernicharo, J., Barlow, M.~J., Gonzalez-Alfonso, E., {et~al.} 1996, A{\&}AL,
  204, L201

\bibitem[{Cesaroni(2005)}]{Cesaroni2005}
Cesaroni, R. 2005, in Proceedings of the International Astronomical Union,
  59--69, \dodoi{10.1017/S1743921305004369}

\bibitem[{Chenel {et~al.}(2016)Chenel, Roncero, Aguado, Ag{\'{u}}ndez, \&
  Cernicharo}]{Chenel2016}
Chenel, A., Roncero, O., Aguado, A., Ag{\'{u}}ndez, M., \& Cernicharo, J. 2016,
  Journal of Chemical Physics, 144, \dodoi{10.1063/1.4945389}

\bibitem[{Clarke {et~al.}(2015)Clarke, Vacca, \& Shuping}]{Clarke2015}
Clarke, M., Vacca, W.~D., \& Shuping, R.~Y. 2015, in Astronomical Data Analysis
  Software and Systems: XXIV, 355--358

\bibitem[{Colzi {et~al.}(2018{\natexlab{a}})Colzi, Fontani, Caselli,
  Ceccarelli, Hily-Blant, \& Bizzocchi}]{Colzi2018b}
Colzi, L., Fontani, F., Caselli, P., {et~al.} 2018{\natexlab{a}}, A{\&}A, 609,
  1, \dodoi{10.1051/0004-6361/201730576}

\bibitem[{Colzi {et~al.}(2018{\natexlab{b}})Colzi, Fontani, Rivilla,
  S{\'{a}}nchez-Monge, Testi, Beltr{\'{a}}n, \& Caselli}]{Colzi2018a}
Colzi, L., Fontani, F., Rivilla, V.~M., {et~al.} 2018{\natexlab{b}}, MNRAS,
  478, 3693, \dodoi{10.1093/MNRAS/STY1027}

\bibitem[{Colzi {et~al.}(2020)Colzi, Sipil{\"{a}}, Roueff, Caselli, \&
  Fontani}]{Colzi2020}
Colzi, L., Sipil{\"{a}}, O., Roueff, E., Caselli, P., \& Fontani, F. 2020,
  A{\&}A.
\newblock \doarXiv{2006.03362}

\bibitem[{Comito {et~al.}(2005)Comito, Schilke, Phillips, Lis, Motte, \&
  Mehringer}]{Comito2005}
Comito, C., Schilke, P., Phillips, T.~G., {et~al.} 2005, ApJS, 156, 127,
  \dodoi{10.1086/425996}

\bibitem[{Crockett {et~al.}(2014)Crockett, Bergin, Neill, Favre, Schilke, Lis,
  Bell, Blake, Cernicharo, Emprechtinger, Esplugues, Gupta, Kleshcheva, Lord,
  Marcelino, McGuire, Pearson, Phillips, Plume, van~der Tak, Tercero, \&
  Yu}]{Crockett2014}
Crockett, N.~R., Bergin, E.~A., Neill, J.~L., {et~al.} 2014, ApJ, 787, 112,
  \dodoi{10.1088/0004-637X/787/2/112}

\bibitem[{Daniel {et~al.}(2013)Daniel, G{\'{e}}rin, Roueff, Cernicharo,
  Marcelino, Lique, Lis, Teyssier, Biver, \&
  Bockel{\'{e}}e-Morvan}]{Daniel2013}
Daniel, F., G{\'{e}}rin, M., Roueff, E., {et~al.} 2013, A{\&}A, 560, 1,
  \dodoi{10.1051/0004-6361/201321939}

\bibitem[{{De Buizer} {et~al.}(2012){De Buizer}, Morris, Becklin, Zinnecker,
  Herter, Adams, Shuping, \& Vacca}]{DeBuizer2012}
{De Buizer}, J.~M., Morris, M.~R., Becklin, E.~E., {et~al.} 2012, ApJ: Letters,
  749, 35, \dodoi{10.1088/2041-8205/749/2/L23}

\bibitem[{{De Graauw} {et~al.}(1996){De Graauw}, Haser, Beintema, Roelfsema,
  {Van Agthoven}, Barl, Bauer, Bekenkamp, Boonstra, Boxhoorn, Cot{\'{e}}, {De
  Groene}, {Van Dijkhuizen}, Drapatz, Evers, Feuchtgruber, Frericks, Genzel,
  Haerendel, Heras, {Van Der Hucht}, {Van Der Hulst}, Huygen, Jacobs, Jakob,
  Kamperman, Katterloher, Kester, Kunze, Kussendrager, Lahuis, Lamers, Leech,
  {Van Der Lei}, {Van Der Linden}, Luinge, Lutz, Melzner, Morris, {Van Nguyen},
  Ploeger, Price, Salama, Schaeidt, Sijm, Smoorenburg, Spakman, Spoon,
  Steinmayer, Stoecker, Valentijn, Vandenbussche, Visser, Waelkens, Waters,
  Wensink, Wesselius, Wiezorrek, Wieprecht, Wijnbergen, Wildeman, \&
  Young}]{deGraauw1996}
{De Graauw}, T., Haser, L.~N., Beintema, D.~A., {et~al.} 1996, A{\&}A, 315, L47

\bibitem[{Dungee {et~al.}(2018)Dungee, Boogert, Dewitt, Montiel, Richter, Barr,
  Blake, Charnley, Indriolo, Karska, Neufeld, Smith, \& Tielens}]{Dungee2018}
Dungee, R., Boogert, A., Dewitt, C.~N., {et~al.} 2018, ApJ: Letters, 868, L10,
  \dodoi{10.3847/2041-8213/aaeda9}

\bibitem[{Dutrey {et~al.}(1997)Dutrey, Guilloteau, \&
  Gu{\'{e}}lin}]{Dutrey1997}
Dutrey, A., Guilloteau, S., \& Gu{\'{e}}lin, M. 1997, A{\&}A, 317, 55

\bibitem[{Evans {et~al.}(1991)Evans, Lacy, \& Carr}]{Evans1991}
Evans, N.~J., Lacy, J.~H., \& Carr, J.~S. 1991, ApJ, 383, 674.
\newblock \url{file:///D:/TA ABRAR terbaru/TA BATA FOAM/jurnal penelitian/BATU
  BATA.pdf}

\bibitem[{Favre {et~al.}(2011)Favre, Despois, Brouillet, Baudry, Combes,
  Wootten, \& Wlodarczak}]{Favre2011}
Favre, C., Despois, D., Brouillet, N., {et~al.} 2011, A{\&}A, 532, 1

\bibitem[{Favre {et~al.}(2014)Favre, Carvajal, Field, J{\o}rgensen, Bisschop,
  Brouillet, Despois, Baudry, Kleiner, Bergin, Crockett, Neill, Margul{\`{e}}s,
  Huet, \& Demaison}]{Favre2014}
Favre, C., Carvajal, M., Field, D., {et~al.} 2014, ApJS, 215,
  \dodoi{10.1088/0067-0049/215/2/25}

\bibitem[{Feng {et~al.}(2015)Feng, Beuther, Henning, Semenov, Palau, \&
  Mills}]{Feng2015}
Feng, S., Beuther, H., Henning, T., {et~al.} 2015, A{\&}A, 590, C1,
  \dodoi{10.1051/0004-6361/201322725e}

\bibitem[{Gao \& Solomon(2004)}]{Gao2004b}
Gao, Y., \& Solomon, P.~M. 2004, ApJS, 152, 63, \dodoi{10.1086/383003}

\bibitem[{Genzel {et~al.}(1981)Genzel, Reid, Moran, \& Downes}]{Genzel1981}
Genzel, R., Reid, M.~J., Moran, J.~M., \& Downes, D. 1981, ApJ, 244, 884,
  \dodoi{10.2307/2034794}

\bibitem[{Giesen {et~al.}(2020)Giesen, Mookerjea, Fuchs, Breier, Witsch, Simon,
  \& Stutzki}]{Giesen2020}
Giesen, T.~F., Mookerjea, B., Fuchs, G.~W., {et~al.} 2020, A{\&}A, 633, A120,
  \dodoi{10.1051/0004-6361/201936538}

\bibitem[{Goddi {et~al.}(2011)Goddi, Greenhill, Humphreys, Chandler, \&
  Matthews}]{Goddi2011}
Goddi, C., Greenhill, L.~J., Humphreys, E.~M., Chandler, C.~J., \& Matthews,
  L.~D. 2011, ApJ: Letters, 739, L13, \dodoi{10.1088/2041-8205/739/1/L13}

\bibitem[{Goldsmith {et~al.}(1986)Goldsmith, Irvine, Hjalmarson, \&
  Ellder}]{Goldsmith1986}
Goldsmith, P.~F., Irvine, W.~M., Hjalmarson, A., \& Ellder, J. 1986, ApJ, 310,
  383, \dodoi{10.1017/CBO9781107415324.004}

\bibitem[{Goldsmith \& Langer(1999)}]{Goldsmith1999}
Goldsmith, P.~F., \& Langer, W.~D. 1999, ApJ, 517, 209, \dodoi{10.1086/307195}

\bibitem[{Goldsmith {et~al.}(1981)Goldsmith, Langer, {Ellder, Joel}, Irvine, \&
  Kollberg}]{Goldsmith1981}
Goldsmith, P.~F., Langer, W.~D., {Ellder, Joel}, Irvine, W., \& Kollberg, E.
  1981, ApJ, 249, 524, \dodoi{10.1017/CBO9781107415324.004}

\bibitem[{G{\'{o}}mez {et~al.}(2008)G{\'{o}}mez, Rodr{\'{i}}guez, Loinard,
  Lizano, Allen, Poveda, \& Menten}]{Gomez2008}
G{\'{o}}mez, L., Rodr{\'{i}}guez, L.~F., Loinard, L., {et~al.} 2008, ApJ, 685,
  333

\bibitem[{Gong {et~al.}(2015)Gong, Henkel, Thorwirth, Spezzano, Menten,
  Walmsley, Wyrowski, Mao, \& Klein}]{Gong2015}
Gong, Y., Henkel, C., Thorwirth, S., {et~al.} 2015, A{\&}A, 581, 1,
  \dodoi{10.1051/0004-6361/201526275}

\bibitem[{Gordon {et~al.}(2017)Gordon, Rothman, Hill, Kochanov, Tan, Bernath,
  Birk, Boudon, Campargue, Chance, Drouin, Flaud, Gamache, Hodges, Jacquemart,
  Perevalov, Perrin, Shine, Smith, Tennyson, Toon, Tran, Tyuterev, Barbe,
  Cs{\'{a}}sz{\'{a}}r, Devi, Furtenbacher, Harrison, Hartmann, Jolly, Johnson,
  Karman, Kleiner, Kyuberis, Loos, Lyulin, Massie, Mikhailenko, Moazzen-Ahmadi,
  M{\"{u}}ller, Naumenko, Nikitin, Polyansky, Rey, Rotger, Sharpe, Sung,
  Starikova, Tashkun, Auwera, Wagner, Wilzewski, Wcis{\l}o, Yu, \&
  Zak}]{Gordon2017}
Gordon, I.~E., Rothman, L.~S., Hill, C., {et~al.} 2017, Journal of Quantitative
  Spectroscopy and Radiative Transfer, 203, 3,
  \dodoi{10.1016/j.jqsrt.2017.06.038}

\bibitem[{Graninger {et~al.}(2015)Graninger, {\"{O}}berg, Qi, \&
  Kastner}]{Graninger2015}
Graninger, D., {\"{O}}berg, K.~I., Qi, C., \& Kastner, J. 2015, ApJ: Letters,
  807, L15, \dodoi{10.1088/2041-8205/807/1/L15}

\bibitem[{Graninger {et~al.}(2014)Graninger, Herbst, {\"{O}}berg, \&
  Vasyunin}]{Graninger2014}
Graninger, D.~M., Herbst, E., {\"{O}}berg, K.~I., \& Vasyunin, A.~I. 2014, ApJ,
  787, 74, \dodoi{10.1088/0004-637X/787/1/74}

\bibitem[{Gu{\'{e}}lin {et~al.}(2007)Gu{\'{e}}lin, Salom{\'{e}}, Neri,
  Garc{\'{i}}a-Burillo, Graci{\'{a}}-Carpio, Cernicharo, Cox, Planesas,
  Solomon, Tacconi, \& Bout}]{Guelin2007}
Gu{\'{e}}lin, M., Salom{\'{e}}, P., Neri, R., {et~al.} 2007, A{\&}A, 462, 13,
  \dodoi{10.1051/0004-6361:20066555}

\bibitem[{Hacar {et~al.}(2020)Hacar, Bosman, \& van Dishoeck}]{Hacar2020}
Hacar, A., Bosman, A.~D., \& van Dishoeck, E.~F. 2020, A{\&}A, 635, A4,
  \dodoi{10.1051/0004-6361/201936516}

\bibitem[{Harada {et~al.}(2015)Harada, Riquelme, Viti, Jim{\'{e}}nez-Serra,
  Requena-Torres, Menten, Mart{\'{i}}n, Aladro, Martin-Pintado, \&
  Hochg{\"{u}}rtel}]{Harada2015}
Harada, N., Riquelme, D., Viti, S., {et~al.} 2015, A{\&}A, 584, A102,
  \dodoi{10.1051/0004-6361/201526994}

\bibitem[{Harris {et~al.}(1995)Harris, Avery, Schuster, Tacconi, \&
  Genzel}]{Harris1995}
Harris, A.~I., Avery, L.~W., Schuster, K.-F., Tacconi, L.~J., \& Genzel, R.
  1995, ApJ, 446, L85, \dodoi{10.1086/187936}

\bibitem[{Harris {et~al.}(2003)Harris, Pavlenko, Jones, \&
  Tennyson}]{Harris2003}
Harris, G.~J., Pavlenko, Y.~V., Jones, H.~R., \& Tennyson, J. 2003, MNRAS, 344,
  1107, \dodoi{10.1046/j.1365-8711.2003.06886.x}

\bibitem[{Harris {et~al.}(2006)Harris, Tennyson, Kaminsky, Pavlenko, \&
  Jones}]{Harris2006}
Harris, G.~J., Tennyson, J., Kaminsky, B.~M., Pavlenko, Y.~V., \& Jones, H.~R.
  2006, Monthly Notices of the Royal Astronomical Society, 367, 400,
  \dodoi{10.1111/j.1365-2966.2005.09960.x}

\bibitem[{Henkel {et~al.}(1988)Henkel, Mauersberger, \& Schilke}]{Henkel1988}
Henkel, C., Mauersberger, R., \& Schilke, P. 1988, A{\&}A, 201, L23

\bibitem[{Herbst(1978)}]{Herbst1978}
Herbst, E. 1978, ApJ, 222, 508, \dodoi{10.1017/CBO9781107415324.004}

\bibitem[{{Hern{\'{a}}ndez Vera} {et~al.}(2017){Hern{\'{a}}ndez Vera}, Lique,
  Dumouchel, Hily-Blant, \& Faure}]{HernandezVera2017}
{Hern{\'{a}}ndez Vera}, M., Lique, F., Dumouchel, F., Hily-Blant, P., \& Faure,
  A. 2017, Monthly Notices of the Royal Astronomical Society, 468, 1084,
  \dodoi{10.1093/mnras/stx422}

\bibitem[{Herpin \& Cernicharo(2000)}]{Herpin2000}
Herpin, F., \& Cernicharo, J. 2000, ApJ, 530, L129, \dodoi{10.1086/312507}

\bibitem[{Hirota {et~al.}(2015)Hirota, Kim, Kurono, \& Honma}]{Hirota2015}
Hirota, T., Kim, M.~K., Kurono, Y., \& Honma, M. 2015, ApJ, 801, 82,
  \dodoi{10.1088/0004-637X/801/2/82}

\bibitem[{Hirota {et~al.}(2017)Hirota, Machida, Matsushita, Motogi, Matsumoto,
  Kim, Burns, \& Honma}]{Hirota2017}
Hirota, T., Machida, M.~N., Matsushita, Y., {et~al.} 2017, Nature Astronomy, 1,
  1, \dodoi{10.1038/s41550-017-0146}

\bibitem[{Hirota {et~al.}(1998)Hirota, Yamamoto, Mikami, \&
  Ohishi}]{Hirota1998}
Hirota, T., Yamamoto, S., Mikami, H., \& Ohishi, M. 1998, ApJ, 503, 717

\bibitem[{Ho {et~al.}(1979)Ho, Barrett, Myers, Matsakis, Cheung, Chui, Towners,
  \& Yngvesson}]{Ho1979}
Ho, P. T.~P., Barrett, A.~H., Myers, P.~C., {et~al.} 1979, ApJ, 234, 912,
  \dodoi{10.1017/CBO9781107415324.004}

\bibitem[{Houck {et~al.}(2004)Houck, Roellig, {Van Cleve}, Forrest, Herter,
  Lawrence, Matthews, Reitsema, Soifer, Watson, Weedman, Huisjen, Troeltzsch,
  Barry, Bernard-Salas, Blacken, Brandl, Charmandaris, Devost, Gull, Hall,
  Henderson, Higdon, Pirger, Schoenwald, Sloan, Uchida, Appleton, Armus,
  Burgdorf, Fajardo-Acosta, Grillmair, Ingalls, Morris, \& Teplitz}]{Houck2004}
Houck, J.~R., Roellig, T.~L., {Van Cleve}, J., {et~al.} 2004, ApJ, 154, 18,
  \dodoi{10.1117/12.550517}

\bibitem[{Hrivnak {et~al.}(2000)Hrivnak, Volk, \& Kwok}]{Hrivnak2000}
Hrivnak, B.~J., Volk, K., \& Kwok, S. 2000, ApJ, 535, 275,
  \dodoi{10.1086/308823}

\bibitem[{Hunter(2007)}]{Hunter2007}
Hunter, J.~D. 2007, Computing in Science and Engineering, 9, 99,
  \dodoi{10.1109/MCSE.2007.55}

\bibitem[{Indriolo {et~al.}(2015{\natexlab{a}})Indriolo, Neufeld, DeWitt,
  Richter, Boogert, Harper, Jaffe, Kulas, McKelvey, Ryde, \&
  Vacca}]{Indriolo2015a}
Indriolo, N., Neufeld, D.~A., DeWitt, C.~N., {et~al.} 2015{\natexlab{a}}, ApJ,
  802, L14, \dodoi{10.1088/2041-8205/802/2/L14}

\bibitem[{Indriolo {et~al.}(2015{\natexlab{b}})Indriolo, Neufeld, Gerin,
  Schilke, Benz, Winkel, Menten, Chambers, Black, Bruderer, Falgarone, Godard,
  Goicoechea, Gupta, Lis, Ossenkopf, Persson, Sonnentrucker, {Van Der Tak},
  {Van Dishoeck}, Wolfire, \& Wyrowski}]{Indriolo2015}
Indriolo, N., Neufeld, D.~A., Gerin, M., {et~al.} 2015{\natexlab{b}}, ApJ, 800,
  40, \dodoi{10.1088/0004-637X/800/1/40}

\bibitem[{Indriolo {et~al.}(2020)Indriolo, Neufeld, Barr, Boogert, DeWitt,
  Karska, Montiel, Richter, \& Tielens}]{Indriolo2020}
Indriolo, N., Neufeld, D.~A., Barr, A.~G., {et~al.} 2020, ApJ, 894, 107,
  \dodoi{10.3847/1538-4357/ab88a1}

\bibitem[{Irvine \& Schloerb(1984)}]{Irvine1984}
Irvine, W.~M., \& Schloerb, F.~P. 1984, ApJ, 282, 516

\bibitem[{Jacquinet-Husson {et~al.}(2016)Jacquinet-Husson, Armante, Scott,
  Ch{\'{e}}din, Cr{\'{e}}peau, Boutammine, Bouhdaoui, Crevoisier, Capelle,
  Boonne, Poulet-Crovisier, Barbe, {Chris Benner}, Boudon, Brown, Buldyreva,
  Campargue, Coudert, Devi, Down, Drouin, Fayt, Fittschen, Flaud, Gamache,
  Harrison, Hill, Hodnebrog, Hu, Jacquemart, Jolly, Jim{\'{e}}nez, Lavrentieva,
  Liu, Lodi, Lyulin, Massie, Mikhailenko, M{\"{u}}ller, Naumenko, Nikitin,
  Nielsen, Orphal, Perevalov, Perrin, Polovtseva, Predoi-Cross, Rotger, Ruth,
  Yu, Sung, Tashkun, Tennyson, Tyuterev, {Vander Auwera}, Voronin, \&
  Makie}]{Jacquinet-Husson2016}
Jacquinet-Husson, N., Armante, R., Scott, N.~A., {et~al.} 2016, Journal of
  Molecular Spectroscopy, 327, 31, \dodoi{10.1016/j.jms.2016.06.007}

\bibitem[{Jin {et~al.}(2015)Jin, Lee, \& Kim}]{Jin2015}
Jin, M., Lee, J.~E., \& Kim, K.~T. 2015, ApJ, 219, 2,
  \dodoi{10.1088/0067-0049/219/1/2}

\bibitem[{J{\o}rgensen {et~al.}(2018)J{\o}rgensen, M{\"{u}}ller, Calcutt,
  Coutens, Drozdovskaya, {\"{O}}berg, Persson, Taquet, {Van Dishoeck}, \&
  Wampfler}]{Jorgensen2018}
J{\o}rgensen, J.~K., M{\"{u}}ller, H.~S., Calcutt, H., {et~al.} 2018, A{\&}A,
  620, A170, \dodoi{10.1051/0004-6361/201731667}

\bibitem[{Kastner {et~al.}(1997)Kastner, Zuckerman, Weintraub, \&
  Forveille}]{Kastner1997}
Kastner, J.~H., Zuckerman, B., Weintraub, D.~A., \& Forveille, T. 1997,
  Science, 277, 67, \dodoi{10.1126/science.277.5322.67}

\bibitem[{{Knez} {et~al.}(2001){Knez}, {Boonman}, {Lacy}, {Evans}, \&
  {Richter}}]{Knez2001}
{Knez}, C., {Boonman}, A.~M.~S., {Lacy}, J.~H., {Evans}, N.~J., I., \&
  {Richter}, M.~J. 2001, in American Astronomical Society Meeting Abstracts,
  Vol. 199, American Astronomical Society Meeting Abstracts, 134.10

\bibitem[{Knez {et~al.}(2009)Knez, Lacy, Evans, {Van Dishoeck}, \&
  Richter}]{Knez2009}
Knez, C., Lacy, J.~H., Evans, N.~J., {Van Dishoeck}, E.~F., \& Richter, M.~J.
  2009, ApJ, 696, 471, \dodoi{10.1088/0004-637X/696/1/471}

\bibitem[{Kochanov {et~al.}(2016)Kochanov, Gordon, Rothman, Wcis{\l}o, Hill, \&
  Wilzewski}]{Kochanov2016}
Kochanov, R.~V., Gordon, I.~E., Rothman, L.~S., {et~al.} 2016, Journal of
  Quantitative Spectroscopy and Radiative Transfer, 177, 15,
  \dodoi{10.1016/j.jqsrt.2016.03.005}

\bibitem[{Kounkel {et~al.}(2017)Kounkel, Hartmann, Loinard, Ortiz-Le{\'{o}}n,
  Mioduszewski, Rodr{\'{i}}guez, Dzib, Torres, Pech, Galli, Rivera, Boden,
  {Evans II}, Brice{\~{n}}o, \& Tobin}]{Kounkel2017}
Kounkel, M., Hartmann, L., Loinard, L., {et~al.} 2017, APJ, 834, 142,
  \dodoi{10.3847/1538-4357/834/2/142}

\bibitem[{{Lacy} {et~al.}(2005){Lacy}, {Knez}, {Evans}, \&
  {Richter}}]{Lacy2005}
{Lacy}, J.~H., {Knez}, C., {Evans}, N.~J., \& {Richter}, M.~J. 2005, in
  American Astronomical Society Meeting Abstracts, Vol. 207, American
  Astronomical Society Meeting Abstracts, 81.22

\bibitem[{Lacy {et~al.}(2002)Lacy, Richter, Greathouse, Jaffe, \&
  Zhu}]{Lacy2002}
Lacy, J.~H., Richter, M.~J., Greathouse, T.~K., Jaffe, D.~T., \& Zhu, Q. 2002,
  PASP, 114, 153, \dodoi{10.1086/338730}

\bibitem[{Lahuis \& van Dishoeck(2000)}]{Lahuis2000}
Lahuis, F., \& van Dishoeck, E.~F. 2000, A{\&}A, 355, 699

\bibitem[{Lahuis {et~al.}(2006)Lahuis, van Dishoeck, Boogert, Pontoppidan,
  Blake, Dullemond, {Evans II}, Hogerheijde, J{\o}rgensen, Kessler-Silacci, \&
  Knez}]{Lahuis2006}
Lahuis, F., van Dishoeck, E.~F., Boogert, A. C.~A., {et~al.} 2006, ApJ, 636,
  L145, \dodoi{10.1086/500084}

\bibitem[{Lee \& Rendell(1991)}]{Lee1991}
Lee, T.~J., \& Rendell, A.~P. 1991, Chemical Physics Letters, 177, 491,
  \dodoi{10.1016/0009-2614(91)90073-I}

\bibitem[{Lis {et~al.}(1997)Lis, Keene, Young, Phillips, Bockel{\'{e}}e-Morvan,
  Crovisier, Schilke, Goldsmith, \& Bergin}]{Lis1997}
Lis, D.~C., Keene, J., Young, K., {et~al.} 1997, Icarus, 130, 355,
  \dodoi{10.1006/icar.1997.5833}

\bibitem[{Liszt \& Lucas(2001)}]{Liszt2001}
Liszt, H., \& Lucas, R. 2001, A{\&}A, 370, 576

\bibitem[{Lo {et~al.}(2015)Lo, Chou, Peng, Lin, Lu, \& Cheng}]{Lo2015}
Lo, J.~I., Chou, S.~L., Peng, Y.~C., {et~al.} 2015, ApJS, 221, 20,
  \dodoi{10.1088/0067-0049/221/1/20}

\bibitem[{Loison {et~al.}(2014)Loison, Wakelam, \& Hickson}]{Loison2014}
Loison, J.~C., Wakelam, V., \& Hickson, K.~M. 2014, MNRAS, 443, 398,
  \dodoi{10.1093/mnras/stu1089}

\bibitem[{Lord(1992)}]{Lord1992}
Lord, S.~D. 1992, {A New Software Tool for Computing Earth's Atmospheric
  Transmission of Near- and Far-Infrared Radiation}.
\newblock \url{https://atran.arc.nasa.gov/cgi-bin/atran/atran.cgi}

\bibitem[{Loughnane {et~al.}(2012)Loughnane, Redman, Thompson, Lo, O'Dwyer, \&
  Cunningham}]{Loughnane2012}
Loughnane, R.~M., Redman, M.~P., Thompson, M.~A., {et~al.} 2012, MNRAS, 420,
  1367, \dodoi{10.1111/j.1365-2966.2011.20121.x}

\bibitem[{Magalh{\~{a}}es {et~al.}(2018)Magalh{\~{a}}es, Hily-Blant, Faure,
  Hernandez-Vera, \& Lique}]{Magalhaes2018}
Magalh{\~{a}}es, V.~S., Hily-Blant, P., Faure, A., Hernandez-Vera, M., \&
  Lique, F. 2018, A{\&}A, 615, 1, \dodoi{10.1051/0004-6361/201832622}

\bibitem[{McGuire(2018)}]{McGuire2018}
McGuire, B.~A. 2018, ApJS, 239, 17, \dodoi{10.3847/1538-4365/aae5d2}

\bibitem[{Mendes {et~al.}(2012)Mendes, Buhr, Berg, Froese, Grieser, Heber,
  Jordon-Thaden, Krantz, Novotn{\'{y}}, Novotny, Orlov, Petrignani, Rappaport,
  Repnow, Schwalm, Shornikov, St{\"{u}}tzel, Zajfman, \& Wolf}]{Mendes2012}
Mendes, M.~B., Buhr, H., Berg, M.~H., {et~al.} 2012, ApJ: Letters, 746, 3,
  \dodoi{10.1088/2041-8205/746/1/L8}

\bibitem[{Miettinen(2014)}]{Miettinen2014}
Miettinen, O. 2014, A{\&}A, 562, 1, \dodoi{10.1051/0004-6361/201322596}

\bibitem[{Milam {et~al.}(2005)Milam, Savage, Brewster, Ziurys, \&
  Wyckoff}]{Milam2005}
Milam, S.~N., Savage, C., Brewster, M.~A., Ziurys, L.~M., \& Wyckoff, S. 2005,
  ApJ, 634, 1126, \dodoi{10.1086/497123}

\bibitem[{Moreno {et~al.}(2011)Moreno, Lellouch, Lara, Courtin,
  Bockel{\'{e}}e-Morvan, Hartogh, Rengel, Biver, Banaszkiewicz, \&
  Gonz{\'{a}}lez}]{Moreno2011}
Moreno, R., Lellouch, E., Lara, L.~M., {et~al.} 2011, A{\&}AL, 536, L12,
  \dodoi{10.1051/0004-6361/201118189}

\bibitem[{Nguyen {et~al.}(2015)Nguyen, Baraban, Ruscic, \&
  Stanton}]{Nguyen2015}
Nguyen, T.~L., Baraban, J.~H., Ruscic, B., \& Stanton, J.~F. 2015, Journal of
  Physical Chemistry A, 119, 10929, \dodoi{10.1021/acs.jpca.5b08406}

\bibitem[{Ohishi(1997)}]{Ohishi1997}
Ohishi, M. 1997, in International Astronomical Union Symposium, 61--74

\bibitem[{Okumura {et~al.}(2011)Okumura, Yamashita, Sako, Miyata, Honda,
  Kataza, \& Okamoto}]{Okumura2011}
Okumura, S.~I., Yamashita, T., Sako, S., {et~al.} 2011, Publ. Astron. Soc.
  Japan, 63, 823, \dodoi{10.1093/pasj/63.4.823}

\bibitem[{Orozco-Aguilera {et~al.}(2017)Orozco-Aguilera, Zapata, Hirota, Qin,
  \& Masqu{\'{e}}}]{Orozco-Aguilera2017}
Orozco-Aguilera, M.~T., Zapata, L.~A., Hirota, T., Qin, S.-L., \& Masqu{\'{e}},
  J.~M. 2017, ApJ, 847, 66, \dodoi{10.3847/1538-4357/aa88cd}

\bibitem[{Peng {et~al.}(2019)Peng, Rivilla, Zhang, Ge, \& Zhou}]{Peng2019}
Peng, Y., Rivilla, V.~M., Zhang, L., Ge, J.~X., \& Zhou, B. 2019, ApJ, 871,
  251, \dodoi{10.3847/1538-4357/aafad4}

\bibitem[{P{\'{e}}rez-Beaupuits {et~al.}(2007)P{\'{e}}rez-Beaupuits, Aalto, \&
  Gerebro}]{Perez-Beaupuits2007}
P{\'{e}}rez-Beaupuits, J.~P., Aalto, S., \& Gerebro, H. 2007, A{\&}A, 476, 177,
  \dodoi{10.1051/0004-6361:20078479}

\bibitem[{Persson {et~al.}(2007)Persson, Olofsson, Koning, Bergman, Bernath,
  Black, Frisk, Geppert, Hasegawa, Hjalmarson, Kwok, Larsson, Lecacheux,
  Nummelin, Olberg, Sandqvist, \& Wirstr{\"{o}}m}]{Persson2007}
Persson, C.~M., Olofsson, A. O.~H., Koning, N., {et~al.} 2007, A{\&}A, 476,
  807, \dodoi{10.1051/0004-6361:20077229}

\bibitem[{Plambeck \& Wright(2016)}]{Plambeck2016}
Plambeck, R.~L., \& Wright, M. C.~H. 2016, ApJ, 833, 219,
  \dodoi{10.3847/1538-4357/833/2/219}

\bibitem[{Plambeck {et~al.}(2009)Plambeck, Wright, Friedel, {Widicus Weaver},
  Bolatto, Pound, Woody, Lamb, \& Scott}]{Plambeck2009}
Plambeck, R.~L., Wright, M.~C., Friedel, D.~N., {et~al.} 2009, ApJ, 704, 28,
  \dodoi{10.1088/0004-637X/704/1/L25}

\bibitem[{Plume {et~al.}(2012)Plume, Bergin, Phillips, Lis, Wang, Crockett,
  Caux, Comito, Goldsmith, \& Schilke}]{Plume2012}
Plume, R., Bergin, E.~A., Phillips, T.~G., {et~al.} 2012, ApJ, 744,
  \dodoi{10.1088/0004-637X/744/1/28}

\bibitem[{Price-Whelan {et~al.}(2018)Price-Whelan, Sipőcz, G{\"{u}}nther, Lim,
  Crawford, Conseil, Shupe, Craig, Dencheva, Ginsburg, VanderPlas, Bradley,
  P{\'{e}}rez-Su{\'{a}}rez, de~Val-Borro, Aldcroft, Cruz, Robitaille, Tollerud,
  Ardelean, Babej, Bach, Bachetti, Bakanov, Bamford, Barentsen, Barmby,
  Baumbach, Berry, Biscani, Boquien, Bostroem, Bouma, Brammer, Bray,
  Breytenbach, Buddelmeijer, Burke, Calderone, Rodr{\'{i}}guez, Cara, Cardoso,
  Cheedella, Copin, Corrales, Crichton, D'Avella, Deil, Depagne, Dietrich,
  Donath, Droettboom, Earl, Erben, Fabbro, Ferreira, Finethy, Fox, Garrison,
  Gibbons, Goldstein, Gommers, Greco, Greenfield, Groener, Grollier, Hagen,
  Hirst, Homeier, Horton, Hosseinzadeh, Hu, Hunkeler, Ivezi{\'{c}}, Jain,
  Jenness, Kanarek, Kendrew, Kern, Kerzendorf, Khvalko, King, Kirkby, Kulkarni,
  Kumar, Lee, Lenz, Littlefair, Ma, Macleod, Mastropietro, McCully, Montagnac,
  Morris, Mueller, Mumford, Muna, Murphy, Nelson, Nguyen, Ninan, N{\"{o}}the,
  Ogaz, Oh, Parejko, Parley, Pascual, Patil, Patil, Plunkett, Prochaska,
  Rastogi, Janga, Sabater, Sakurikar, Seifert, Sherbert, Sherwood-Taylor, Shih,
  Sick, Silbiger, Singanamalla, Singer, Sladen, Sooley, Sornarajah, Streicher,
  Teuben, Thomas, Tremblay, Turner, Terr{\'{o}}n, van Kerkwijk, de~la Vega,
  Watkins, Weaver, Whitmore, Woillez, \& Zabalza}]{Price-Whelan2018}
Price-Whelan, A.~M., Sipőcz, B.~M., G{\"{u}}nther, H.~M., {et~al.} 2018, ApJ,
  156, 123, \dodoi{10.3847/1538-3881/aabc4f}

\bibitem[{Rangwala {et~al.}(2018)Rangwala, Colgan, {Le Gal}, Acharyya, Huang,
  Lee, Herbst, Richter, Boogert, \& Mckelvey}]{Rangwala2018}
Rangwala, N., Colgan, S. W.~J., {Le Gal}, R., {et~al.} 2018, ApJ, 856, 9,
  \dodoi{10.3847/1538-4357/aaab66}

\bibitem[{Richter {et~al.}(2018)Richter, DeWitt, McKelvey, Montiel, McMurray,
  \& Case}]{Richter2018}
Richter, M.~J., DeWitt, C.~N., McKelvey, M., {et~al.} 2018, Journal of
  Astronomical Instrumentation, 7, 1, \dodoi{10.1142/S2251171718400135}

\bibitem[{Rickard {et~al.}(1977)Rickard, Palmer, Turner, Morris, \&
  Zuckerman}]{Rickard1977}
Rickard, L.~J., Palmer, P., Turner, B.~E., Morris, M., \& Zuckerman, B. 1977,
  ApJ, 214, 390

\bibitem[{Rivilla {et~al.}(2017)Rivilla, Beltr{\'{a}}n, Cesaroni, Fontani,
  Codella, \& Zhang}]{Rivilla2017}
Rivilla, V.~M., Beltr{\'{a}}n, M.~T., Cesaroni, R., {et~al.} 2017, A{\&}A, 598,
  1, \dodoi{10.1051/0004-6361/201628373}

\bibitem[{Robitaille {et~al.}(2013)Robitaille, Tollerud, Greenfield,
  Droettboom, Bray, Aldcroft, Davis, Ginsburg, Price-Whelan, Kerzendorf,
  Conley, Crighton, Barbary, Muna, Ferguson, Grollier, Parikh, Nair,
  G{\"{u}}nther, Deil, Woillez, Conseil, Kramer, Turner, Singer, Fox, Weaver,
  Zabalza, Edwards, {Azalee Bostroem}, Burke, Casey, Crawford, Dencheva, Ely,
  Jenness, Labrie, Lim, Pierfederici, Pontzen, Ptak, Refsdal, Servillat, \&
  Streicher}]{Robitaille2013}
Robitaille, T.~P., Tollerud, E.~J., Greenfield, P., {et~al.} 2013, A{\&}A, 558,
  1, \dodoi{10.1051/0004-6361/201322068}

\bibitem[{Rolffs {et~al.}(2011)Rolffs, Schilke, Wyrowski, Dullemond, Menten,
  Thorwirth, \& Belloche}]{Rolffs2011}
Rolffs, R., Schilke, P., Wyrowski, F., {et~al.} 2011, A{\&}A, 529, 1,
  \dodoi{10.1051/0004-6361/201116544}

\bibitem[{Sarrasin {et~al.}(2010)Sarrasin, {Ben Abdallah}, Wernli, Faure,
  Cernicharo, \& Lique}]{Sarrasin2010}
Sarrasin, E., {Ben Abdallah}, D., Wernli, M., {et~al.} 2010, Monthly Notices of
  the Royal Astronomical Society, 404, 518,
  \dodoi{10.1111/j.1365-2966.2010.16312.x}

\bibitem[{Schilke {et~al.}(2001)Schilke, Benford, Hunter, Lis, \&
  Phillips}]{Schilke2001}
Schilke, P., Benford, D.~J., Hunter, T.~R., Lis, D.~C., \& Phillips, T.~G.
  2001, ApJS, 132, 281, \dodoi{10.1086/318951}

\bibitem[{Schilke {et~al.}(1997)Schilke, Groesbeck, Blake, \&
  Phillips}]{Schilke1997}
Schilke, P., Groesbeck, T.~D., Blake, G.~A., \& Phillips, T.~G. 1997, ApJS,
  108, 301, \dodoi{10.1086/312948}

\bibitem[{Schilke {et~al.}(1992)Schilke, Walmsley, {G. Pineau des
  For{\^{e}}ts}, Roueff, Flower, \& Guilloteau}]{Schilke1992}
Schilke, P., Walmsley, C.~M., {G. Pineau des For{\^{e}}ts}, {et~al.} 1992,
  A{\&}A, 256, 595

\bibitem[{Shuping {et~al.}(2004)Shuping, Morris, \& Bally}]{Shuping2004}
Shuping, R.~Y., Morris, M., \& Bally, J. 2004, AJ, 128, 363,
  \dodoi{10.1086/421373}

\bibitem[{Snyder \& Buhl(1971)}]{Snyder1971}
Snyder, L.~E., \& Buhl, D. 1971, ANYAS, 194, 17

\bibitem[{Snyder \& Buhl(1972)}]{Snyder1972}
---. 1972, ApJ, 177, 619, \dodoi{10.1086/151739}

\bibitem[{Stutzki {et~al.}(1988)Stutzki, Genzel, Harris, Herman, \&
  Jaffe}]{Stutzki1988}
Stutzki, J., Genzel, R., Harris, A.~I., Herman, J., \& Jaffe, D.~T. 1988, ApJ,
  330, L125, \dodoi{10.1086/185219}

\bibitem[{Sutton {et~al.}(1995)Sutton, Peng, Danchi, Jaminet, Sandell, \&
  Russell}]{Sutton1995}
Sutton, E.~C., Peng, R., Danchi, W.~C., {et~al.} 1995, ApJS, 97, 455

\bibitem[{Tennekes {et~al.}(2006)Tennekes, Harju, Juvela, \&
  T{\'{o}}th}]{Tennekes2006}
Tennekes, P.~P., Harju, J., Juvela, M., \& T{\'{o}}th, L.~V. 2006, A{\&}A, 456,
  1037, \dodoi{10.1051/0004-6361:20040294}

\bibitem[{Tercero {et~al.}(2010)Tercero, Cernicharo, Pardo, \&
  Goicoechea}]{Tercero2010}
Tercero, B., Cernicharo, J., Pardo, J.~R., \& Goicoechea, J.~R. 2010, A{\&}A,
  517, \dodoi{10.1051/0004-6361/200913501}

\bibitem[{Turner {et~al.}(1997)Turner, Pirgov, \& Minh}]{Turner1997}
Turner, B.~E., Pirgov, L., \& Minh, Y.~C. 1997, ApJ, 483, 235

\bibitem[{van~der Tak(2004)}]{VanDerTak2004}
van~der Tak, F.~F. 2004, in Star Formation at High Angular Resolution, IAU
  Symposium, 59--66

\bibitem[{van~der Walt {et~al.}(2011)van~der Walt, Colbert, \&
  Varoquaux}]{VanDerWalt2011}
van~der Walt, S., Colbert, S.~C., \& Varoquaux, G. 2011, Computing in Science
  and Engineering, 13, 22, \dodoi{10.1109/MCSE.2011.37}

\bibitem[{van Dishoeck {et~al.}(1998)van Dishoeck, Wright, Helmich, Cernicharo,
  Gonz{\'{a}}lez-Alfonso, {De Graauw}, \& Vandenbussche}]{VanDishoeck1998}
van Dishoeck, E.~F., Wright, C.~M., Helmich, F.~P., {et~al.} 1998, ApJ, 502,
  1996

\bibitem[{Vasyunina {et~al.}(2011)Vasyunina, Linz, Henning, Zinchenko, Beuther,
  \& Voronkov}]{Vasyunina2011}
Vasyunina, T., Linz, H., Henning, T., {et~al.} 2011, A{\&}A, 527, A88,
  \dodoi{10.1051/0004-6361/201014974}

\bibitem[{Virtanen {et~al.}(2020)Virtanen, Gommers, Oliphant, Haberland, Reddy,
  Cournapeau, Burovski, Peterson, Weckesser, Bright, van~der Walt, Brett,
  Wilson, Millman, Mayorov, Nelson, Jones, Kern, Larson, Carey, Polat, Feng,
  Moore, VanderPlas, Laxalde, Perktold, Cimrman, Henriksen, Quintero, Harris,
  Archibald, Ribeiro, Pedregosa, van Mulbregt, Vijaykumar, Bardelli, Rothberg,
  Hilboll, Kloeckner, Scopatz, Lee, Rokem, Woods, Fulton, Masson,
  H{\"{a}}ggstr{\"{o}}m, Fitzgerald, Nicholson, Hagen, Pasechnik, Olivetti,
  Martin, Wieser, Silva, Lenders, Wilhelm, Young, Price, Ingold, Allen, Lee,
  Audren, Probst, Dietrich, Silterra, Webber, Slavi{\v{c}}, Nothman, Buchner,
  Kulick, Sch{\"{o}}nberger, {de Miranda Cardoso}, Reimer, Harrington,
  Rodr{\'{i}}guez, Nunez-Iglesias, Kuczynski, Tritz, Thoma, Newville,
  K{\"{u}}mmerer, Bolingbroke, Tartre, Pak, Smith, Nowaczyk, Shebanov, Pavlyk,
  Brodtkorb, Lee, McGibbon, Feldbauer, Lewis, Tygier, Sievert, Vigna, Peterson,
  More, Pudlik, Oshima, Pingel, Robitaille, Spura, Jones, Cera, Leslie, Zito,
  Krauss, Upadhyay, Halchenko, \& V{\'{a}}zquez-Baeza}]{Virtanen2020}
Virtanen, P., Gommers, R., Oliphant, T.~E., {et~al.} 2020, Nature Methods, 17,
  261, \dodoi{10.1038/s41592-019-0686-2}

\bibitem[{Wampfler {et~al.}(2014)Wampfler, J{\o}rgensen, Bizzarro, \&
  Bisschop}]{Wampfler2014}
Wampfler, S.~F., J{\o}rgensen, J.~K., Bizzarro, M., \& Bisschop, S.~E. 2014,
  A{\&}A, 572, 1, \dodoi{10.1051/0004-6361/201423773}

\bibitem[{Willis {et~al.}(2020)Willis, Garrod, Belloche, M{\"{u}}ller, Barger,
  Bonfand, \& Menten}]{Willis2020}
Willis, E.~R., Garrod, R.~T., Belloche, A., {et~al.} 2020, A{\&}A, 636, 1,
  \dodoi{10.1051/0004-6361/201936489}

\bibitem[{Wright \& Plambeck(2017)}]{Wright2017}
Wright, M. C.~H., \& Plambeck, R.~L. 2017, ApJ, 843, 83,
  \dodoi{10.3847/1538-4357/aa72e6}

\bibitem[{Wright {et~al.}(1995)Wright, Plambeck, Mundy, \& Looney}]{Wright1995}
Wright, M. C.~H., Plambeck, R.~L., Mundy, L.~G., \& Looney, L.~W. 1995, ApJ,
  455, \dodoi{10.1086/309829}

\bibitem[{Yan {et~al.}(2019)Yan, Zhang, Henkel, Mufakharov, Jia, Tang, Wu, Li,
  Zeng, Wang, Li, Huang, \& Jian}]{Yan2019}
Yan, Y.~T., Zhang, J.~S., Henkel, C., {et~al.} 2019, ApJ, 877, 154,
  \dodoi{10.3847/1538-4357/ab17d6}

\bibitem[{Young {et~al.}(2012)Young, Becklin, Marcum, Roellig, {De Buizer},
  Herter, G{\"{u}}sten, Dunham, Temi, Andersson, Backman, Burgdorf, Caroff,
  Casey, Davidson, Erickson, Gehrz, Harper, Harvey, Helton, Horner, Howard,
  Klein, Krabbe, McLean, Meyer, Miles, Morris, Reach, Rho, Richter, Roeser,
  Sandell, Sankrit, Savage, Smith, Shuping, Vacca, Vaillancourt, Wolf, \&
  Zinnecker}]{Young2012}
Young, E.~T., Becklin, E.~E., Marcum, P.~M., {et~al.} 2012, ApJ: Letters, 749,
  5, \dodoi{10.1088/2041-8205/749/2/L17}

\bibitem[{Zapata {et~al.}(2012)Zapata, Rodr{\'{i}}guez, Schmid-Burgk, Loinard,
  Menten, \& Curiel}]{Zapata2012}
Zapata, L.~A., Rodr{\'{i}}guez, L.~F., Schmid-Burgk, J., {et~al.} 2012, ApJ,
  754, \dodoi{10.1088/2041-8205/754/1/L17}

\bibitem[{Zapata {et~al.}(2011)Zapata, Schmid-Burgk, \& Menten}]{Zapata2011}
Zapata, L.~A., Schmid-Burgk, J., \& Menten, K.~M. 2011, Astronomy and
  Astrophysics, 529, 1, \dodoi{10.1051/0004-6361/201014423}

\bibitem[{Zeng {et~al.}(2017)Zeng, Jim{\'{e}}nez-Serra, Cosentino, Viti,
  Barnes, Henshaw, Caselli, Fontani, \& Hily-Blant}]{Zeng2017}
Zeng, S., Jim{\'{e}}nez-Serra, I., Cosentino, G., {et~al.} 2017, A{\&}A, 603,
  1, \dodoi{10.1051/0004-6361/201630210}

\bibitem[{Zhang {et~al.}(2020)Zhang, Quan, Chang, Herbst, Esimbek, \&
  Webb}]{Zhang2020}
Zhang, X., Quan, D., Chang, Q., {et~al.} 2020, MNRAS, 497, 609,
  \dodoi{10.1093/mnras/staa1979}

\bibitem[{Zuckerman {et~al.}(1972)Zuckerman, Morris, Palmer, \&
  Turner}]{Zuckerman1972}
Zuckerman, B., Morris, M., Palmer, P., \& Turner, B.~E. 1972, APJ, 173, L125

\end{thebibliography}
\bibliographystyle{aasjournal}

\end{document}